\documentclass[iop]{emulateapj}

\usepackage{graphicx,psfig,fancyhdr,natbib,subfigure}
\usepackage{epsfig, psfig, epsf}
\usepackage{amsmath, cancel}
\usepackage{amssymb,float}
\usepackage{longtable, appendix}

 \def\apjl{ApJ~Lett.} 
\def\apjs{ApJS}

 \def\aap{Astron.~\&~Astrophys.}

\def \hmpc{~h^{-1}~{\rm Mpc}} 

\newcommand{\degree}{\ensuremath{^\circ}}

\shorttitle{NIR properties of Quasars}
\shortauthors{Peth, Ross, Schneider} 

\begin{document}

\title{Near Infrared Photometric properties of 130,000 Quasars: \\
         An SDSS-UKIDSS matched catalog}
\author{Michael A. Peth\altaffilmark{1}, 
            Nicholas P. Ross\altaffilmark{1,2}, 
            Donald P. Schneider\altaffilmark{1}}
\altaffiltext{1}{Department of Astronomy and Astrophysics, The Pennsylvania State University, 525 Davey Laboratory, University Park, PA 16802, U.S.A.}
\altaffiltext{2}{Lawrence Berkeley National Lab, 1 Cyclotron Road, Berkeley, CA, 94720, U.S.A. Email: npross@lbl.gov}

\begin{abstract}
We present a catalog of over 130,000 quasars candidates with NIR
photometric properties, with an areal coverage of
approximately \hbox{1,200}~deg$^{2}$. This is achieved by matching the
Sloan Digital Sky Survey (SDSS) in the optical {\it ugriz} bands, to
the UKIRT Infrared Digital Sky Survey (UKIDSS) Large Area Survey (LAS)
in the near-infrared {\it YJHK} bands.  We match the $\approx 1$
million SDSS DR6 Photometric Quasar catalog to Data Release 3 of the
UKIDSS LAS (ULAS), and produce a catalog with \hbox{130,827} objects
with detections in one or more NIR bands, of which \hbox{74,351}
objects have optical and {\it K}-band detections and \hbox{42,133}
objects have the full 9-band photometry. The majority ($\sim85\%$) of
the SDSS objects were not matched simply because there were not
covered by the ULAS.
The positional standard deviation of the SDSS Quasar to ULAS matches
is $\delta_{\rm R.A.} = 0.1370''$ and $\delta_{\rm Decl.}= 0.1314''$.
We find an absolute systematic astrometric offset between the SDSS
Quasar catalog and the UKIDSS LAS, of $|{\rm R.A._{offset}}| = 0.025''$,
and $|{\rm Decl._{offset}}| = 0.040''$; we suggest the nature of this
offset to be due to the matching of catalog, rather than image, level
data.
Our matched catalog has a surface density of $\approx53$ deg$^{-2}$
for $K\leq18.27$ objects; tests using our matched catalog, along
with data from the UKIDSS DXS, implies that our limiting magnitude is
$i\approx20.6$.  Color-redshift diagrams, for the optical and NIR, show the close
agreement between our matched catalog and recent quasar color models
at redshift $z\lesssim2.0$, while at higher redshifts, the models
generally appear to be {\it bluer} than the mean observed quasar
colors.  The $gJK$ and $giK$ color-spaces are used to examine methods of
differentiating between stars and (mid-redshift) quasars, key to
currently ongoing quasar surveys. Finally, we report on the NIR
photometric properties of high, $z>4.6$, and very high, $z>5.7$,
redshift previously discovered quasars.

\end{abstract}


\section{Introduction}  
With the completion of the Sloan Digital Sky Survey
\citep[SDSS;][]{York00}, the era of high quality, homogeneous CCD
imaging over large fractions of the sky has arrived. The final data
release from the SDSS, \citep[DR7;][]{Abazajian09} contains over
11,000 deg$^{2}$ of imaging data, with 357 million unique objects
being identified in a $\sim 60$ Terabyte database.

\begin{table*}
  \begin{center}
    \caption{Descriptions of previous catalogs of quasars. 
             $^{a}$O/I, Optical or Infrared; P/S, Photometric or spectroscopic survey. 
             $^{b}$SDSS Faint Quasar Survey. 
             $^{c}$The FIRST-2MASS Red Quasar Survey.
             $^{d}$The SDSS DR7 Quasar catalog provides NIR photometry from 2MASS for \hbox{53,584} objects, 
                   \hbox{29,551} or which are detected in the {\it K}-band.
             $^{e}$The 2MASS Second Incremental Data Release. }
  \setlength{\tabcolsep}{4pt}
    \begin{tabular}{lrrccl}
      \hline
      \hline
      Survey  & Area (deg$^{2}$) & N$_{\rm Q}$ & Magnitude Range & O/I \& P/S$^{a}$ & Reference  \\
      \hline
      COMBO-17        &            0.8 & \hbox{      192} & $R<24$                & O/P & \citet{Wolf03} \\
      SFQS$^{b}$           &             4  & \hbox{      414} & $g<22.5$              & O/S & \citet{Jiang06} \\
      SDSS-ULAS DR1        &           189 & \hbox{    2 873} & $K<18.2$               &  I/S & \citet{Chiu07}       \\
      2SLAQ QSO           &            190 & \hbox{    8 764} & $18.00 < g < 21.85$   & O/S & \citet{Croom09a}     \\
      2QZ                 &                   700 & \hbox{   23 338} & $18.25<b_{\rm J}<20.85$ & O/S & \citet{Croom04}     \\
      SDSS-ULAS DR3             &         1 200 & \hbox{   74 351} & $K<18.4$               & I/P/S & this work           \\
      FIRST-2MASS RQS$^{c}$ &             2 716 &   \hbox{ 57}     & $K \leq 14.3, (R-K)>4, (J-K)>1.7$ & I/S  & \citet{Glikman07} \\
      SDSS DR6pQ                  &              8 342 & \hbox{1 015 082} & $i<21.3$               & O/P & \citet{Richards09}  \\
      SDSS DR7Q$^{d}$           &              9 380 & \hbox{   29 551} & $K \lesssim 17.0$     & I/S & Schneider et al. 2010\\
      SDSS DR7Q                    &              9 380 & \hbox{  105 807} & $i<20.2$             & O/S & Schneider et al. 2010 \\
      2MASS 2IDR$^{e}$          & $\sim30$ 000 & \hbox{2 277}     & $K \leq 15$            & I/S  & \citet{Barkhouse01} \\
      \hline
      \hline
      \label{tab:previous_surveys}
    \end{tabular}
 \end{center}
\end{table*}

The identification of quasars was a major focus of the SDSS project;
utilizing the broad 5-filter photometry \citep{Richards02} led to
efficient selection \citep{Vanden_Berk05, Richards06} of low, $z<2.2$,
and high, $z\gtrsim 3.5$, redshift quasars, identified via their
spectroscopic signatures \citep[][ and references
therein]{Schneider10}.  However, quasars can also be efficiently
identified by their SDSS imaging properties alone \citep{Richards01},
due to their point-source appearance, but non-stellar location in
color-color space. With the most recent catalog of \citet[][hereafter
R09]{Richards09}, based on imaging from the SDSS, the number of {\it
photometrically identified} quasars now stands at over 1 million. Both
the SDSS spectroscopic and photometric quasar catalogs have been used
to investigate global quasar properties such as the luminosity
function \citep[QLF; ][]{Fan01, Richards06, Croom09b} and clustering
\citep{Myers06, Myers07a, Shen07, Shen09, Ross09}.

The emergence of large surveys has not been confined to the optical
regime. In the near-infrared (NIR; $\lambda \approx 1-5\mu$m)
quasar catalogs \citep[e.g.][]{Barkhouse01, Cutri02, Francis04, Ofek07,
Kouzuma10a, Kouzuma10b} have been available since the completion of
the 2 Micron All Sky Survey \citep[2MASS; ][]{Skrutskie06}.
Table~\ref{tab:previous_surveys} summarizes how previous quasar
surveys, both in the optical and NIR, compare by area, number of
objects and magnitude range.

Quasar observations in the NIR are particularly important for
individual objects that are seen only in the reddest, or indeed
potentially none, of the optical filters \citep[e.g. ][]{Fan06,
Venemans07, Willott09}. Observations of the general quasar population
in the {\it K}-band are key since this links the rest-frame
ultraviolet (UV)/optical to the mid-infrared (MIR; $\lambda \approx
5-30\mu$m); the former being where there is the peak in the radiative
output for Type I, non-obscured quasars, \citep[e.g. ][]{SS73,
Sanders89, Kishimoto08, RIchards09b}, and the latter where reprocessed
light heats intrinsic dust to $\sim 30-300$ K \citep[e.g. ][]{Pier93,
Efstathiou95, Lacy04}.  Observations in the observed {\it K}-band also
measures the rest-frame {\it i} and {\it g}-bands at redshifts
$z\sim1.9$ and $z\sim3.7$, respectively.

However, it is only the bright, $g\lesssim16$ magnitude quasars that
are detected by the relatively shallow limits of the 2MASS survey, and
the majority of known quasars are fainter than this in the NIR
bands. The UKIRT Infrared Deep Sky Survey \citep[UKIDSS;
][]{Lawrence07}, a seven-year sky survey which began in 2005 May, has
five different survey components. The UKIDSS ``Large Area Survey''
(ULAS) aims to reach $\sim 4$ magnitudes deeper than 2MASS over an
area of up to 4,000 deg$^{2}$, directly overlapping the optical
imaging footprint of the SDSS.

The primary goal of this paper is to create a catalog of over 130,000
quasars with optical, {\it ugriz} \citep{Fukugita96}, and NIR, {\it
YJHK} photometry, with an areal coverage of \hbox{1 200}~deg$^{2}$.
By matching the catalogs of R09 in the optical regime to that of the
ULAS in the NIR we produce, by a factor of at least two, the largest
catalog of quasars detected in the {\it K}-band. The major motivation
for the catalog will be its utilization in an future study where we
measure the $K$-band quasar luminosity function (Peth, Ross et al. in
prep.).  By concentrating on the $\approx 200$~deg$^{2}$ area from the
SDSS known as ``Stripe 82'', our {\it K}-band quasar luminosity
function will show the evolution from redshift of zero to two.

The analysis by \citet{Trammell07} is an example of the synergy
produced by the matching and production of a multi-wavelength catalog.
These authors match $\sim$6000 SDSS quasars to UV data provided by the
{\it Galaxy Evolution Explorer} \citep[{\it GALEX}; ][]{Martin05}
satellite and find that over 80$\%$ of the optically detected quasars
have near-UV detections.  The quasars are well separated from stars in
UV-optical color-color space.  The large sample size allows for the
construction of SEDs in bins of redshift and luminosity, which shows
the median SED becoming bluer at UV wavelengths for quasars with lower
continuum luminosity. \citet{Ball07} also perform catalog matching
using SDSS quasar data and {\it GALEX} UV photometry, with the goal of
understanding quasar photometric redshift properties.

The studies by \citet{Warren00}, \citet{Croom01KX}, \citet{Sharp02}
and more recently, \citet{Maddox08} and \citet{Smail08}, are another
motivation why a sample of quasars with NIR photometric properties is
desired.  These authors show that using a ``KX selection'', where the
quasar SED shows an excess in the {\it K}-band compared to a stellar
SED, can successfully identify quasar candidate objects that would be
normally excluded from the SDSS (optical) quasar selection algorithm -
even for dust reddened quasars. This is an important result since
selecting complete quasar samples via the KX method opens up the
possibility of investigating the ``Quasar Epoch'' over the redshift
range of $2.2<z<3.5$, where current, usually optically selected,
quasar samples are particularly poorly represented. A similar project
to \citet{Maddox08} was \citet{Nakos09}, who also select quasar
candidates using the KX-technique, where these authors identify
quasars on the basis of their optical ($R$ and $z^{'}$) to NIR
($K_{\rm s}$) photometry and point-like morphology.  \citet{Jurek08}
also test the KX method and find that it is more effective than the
traditional ``UV Excess'' (UVX) selection method at finding red,
($b_{\rm J} - K) \geq 3.5$, quasars.

There are two comparable studies and samples to our own
work. \citet{Chiu07} and \citet{Souchay09}, the latter recently
producing the ``Large Quasar Astrometric Catalog'' (LQAC).  Our work
differs from these studies, in two key ways; ({\it i}) we have over an
order of magnitude more objects in our sample compared to
\citet{Chiu07}; ({\it ii}) our catalog uses UKIDSS data, as opposed to
2MASS data \citep{Souchay09}, where the former is much better matched
to the SDSS imaging depth.

This paper is organized as follows. In Section 2 we present our
sample, giving a brief overview of the SDSS and UKIDSS.  Section 3
lays out our catalog. In Section 4 we present $N(z)$, color-redshift
and color-color relations from our matched catalog.  In Section 5 we
present analysis of high-redshift quasars and calculations of the {\it
i} and {\it K}-band number counts. The Appendix gives further details
on magnitude conversions and cross-checks of our study.

We assume the currently preferred flat, ``Lambda Cold Dark Matter''
($\Lambda$CDM) cosmology where $\Omega_{\rm b}$ =0.042, $\Omega_{\rm
m}$ = 0.237, $\Omega_{\Lambda}$ = 0.763 \citep{Sanchez06, Spergel07}
and quote distances in units of $\hmpc$ to aid in ease of comparisons
with previous results in the literature.  Where a value of Hubble's
Constant is assumed, e.g. for absolute magnitudes, this will be quoted
explicitly. All optical magnitudes are based and quoted on the AB
zero-point system \citep{Oke83}, while all NIR magnitudes are based in
the {\it Vega} system, with conversions from AB to Vega given in the
Appendix.

\section{Data and Methods}
In this section, we provide overviews of the surveys and the catalogs
we utilize, and how matching was performed.

    \subsection{The SDSS DR6 Photometric Quasar Catalog}
    Details regarding the SDSS can be found in the series of SDSS 
    Data Release papers \citep[and references therein]{Abazajian09}. 
    Full details for the SDSS DR6 Photometric Quasar Catalog (DR6pQ)
    are given in R09, with \citet{Richards04} providing the
    ``proof-of-concept'' study and \citet{Weinstein04} providing an
    empirical algorithm for obtaining photometric redshifts.  Here we
    present the details specific to our work.
    
    The photometric imaging data for the DR6pQ is based upon the SDSS
    Data Release 6 \citep{Adelman-McCarthy08}.  Points sources with PSF
    $i$-band magnitudes between 14.5 and 21.3 are extracted from the SDSS
    Catalog Archive Server (CAS).  We continue the convention of R09,
    utilizing \"{u}bercalibrated magnitudes \citep{Padmanabhan08a} which
    are available in the SDSS database. The \"{u}bercalibrated magnitudes
    represent the most robust photometric measurements as they are
    calibrated across SDSS ``stripes'' to a single uniform photometric
    system for the entire SDSS area. All magnitudes have been corrected
    for Galactic extinction using the \citet{Schlegel98} dust maps.
    
    There are \hbox{1 015 082} objects in total in the R09 catalog
    across \hbox{8 342} deg$^{2}$. The DR6 primary imaging data cover an
    area of 8417 deg$^{2}$, although as noted in R09, due to cuts the
    total effective area covered by this catalog is reduced by $\sim$75
    deg$^{2}$.

    \subsection{The UKIRT Infrared Deep Sky Survey}
    \citet{Lawrence07} gives the general overview for UKIDSS.  In
    brief, the UKIDSS is a collection of five surveys of different
    coverage and depth and using WFCAM \citep{Casali07} on UKIRT.  WFCAM
    has an instantaneous field of view of 0.21 deg$^{2}$, and the various
    surveys employ up to five filters, {\it ZYJHK}, covering the
    wavelength range 0.83-2.37$\mu$m.  The photometric system and
    calibration are described in \citet{Hewett06} and \citet{Hodgkin09},
    respectively. The pipeline processing is described in Irwin et
    al. (2010, in prep.) and the WFCAM Science Archive (WSA) by
    \citet{Hambly08}.  The processed right ascension and declination data
    are accurate to 0.1 arcsec. We have used data from the worldwide 4th
    data release, DR3, which is described in detail by Warren et
    al. (2010, in prep.).  For this paper, we concentrate on the Large
    Area Survey and the Deep Extragalactic Survey.
    
    \begin{figure*}
      \includegraphics[height=22.0cm,width=16.0cm]
      {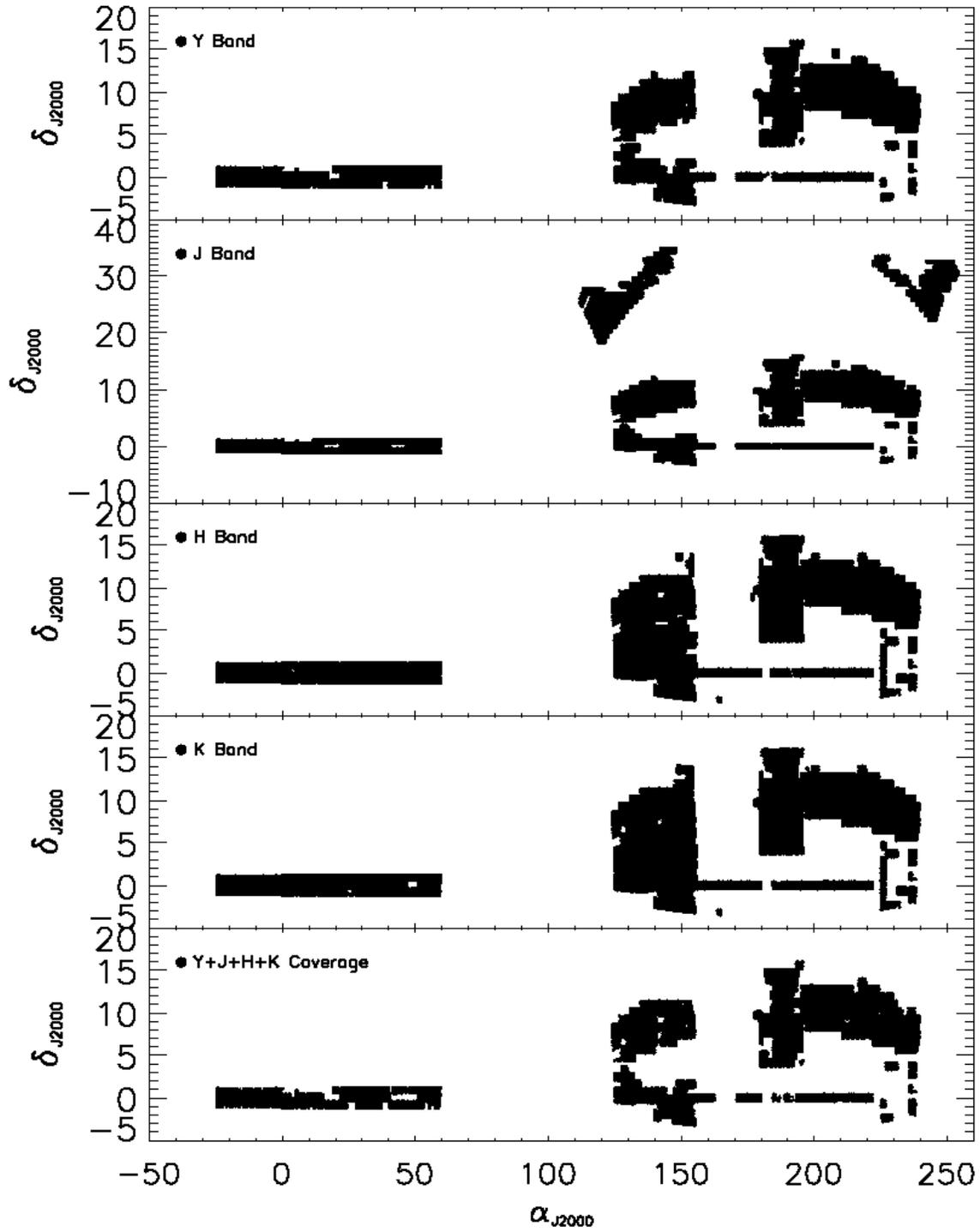} 
      \centering
      \caption[]
      {Coverage over the entire UKIDSS area in each of the 4
       NIR bands and finally in all 4 NIR bands simultaneously.  
       Uses no projection, with axes in degrees
       for both Right Ascension and Declination.  Displays $YJHK$ coverage in
       descending order, followed by coverage in all four bands.}
      \label{fig:fullcov}
    \end{figure*}
    
    \subsubsection{UKIDSS LAS}
    The ULAS aims to map a large fraction of the Northern Sky, $\sim
    4000$ deg$^{2}$, which, when combined with the SDSS, produces an atlas
    covering almost an octave in wavelength. The target depths of the
    survey are $Y=20.3, J=19.5, H=18.6, K = 18.2$ (Vega), and the ULAS
    does not image in the WFCAM $Z$-band.  The ULAS data for our matched
    catalog came courtesy of private communications with Mike Read from
    the WSA. Unlike the SDSS, the ULAS multiband photometry is not taken
    in one observation \citep[e.g. Sec. 5.2 of ][ Sec. 4.2]{Dye06,
      Lawrence07}. Therefore the four bands can, and do, have different
    coverage maps, with the {\it H} and {\it K} bands obtained together,
    and {\it Y} and {\it J} done separately. The ULAS DR3 coverage is
    903 deg$^{2}$, 1,161 deg$^{2}$, 1,091 deg$^{2}$ and 1,111 deg$^{2}$
    in {\it Y}, {\it J}, {\it H} and {\it K} respectively. Note that the
    ultimate 7-year ULAS goal is to cover 4028 deg$^{2}$, in each filter,
    and have two passes of the entire ULAS area with the {\it J}
    filter. The DR3 coverage of each of the four bands, and for all four
    bands, is shown in Figure~\ref{fig:fullcov}. The joint coverage of all four
    bands is 801 deg$^{2}$.

    \subsubsection{UKIDSS DXS} 
    The UKIDSS Deep Extragalactic Survey (DXS) plans to cover 35
    deg$^{2}$ of sky to a 5$\sigma$ point-source sensitivity of {\it
      J}=22.3 and {\it K}=20.8 in four specifically selected multiwavelength
    survey areas. The locations and areas of these four fields, XMM-LSS,
    Lockman Hole, ELAIS N1 and SA22 (a.k.a VIMOS-4), are given in Table 5
    of \citet{Lawrence07}.  Three of these, the Lockman Hole, ELAIS N1 and
    SA22 fields, are covered by the R09 DR6pQ catalog. We shall therefore
    use the DXS data to test the faint end and limiting magnitude of the
    optical and NIR matched quasar catalogs.
    
    \subsection{SDSS Stripe 82} 
    Stripe 82 is a 300 deg$^{2}$ area of repeat photometry on a 2.5
    degree wide stripe, centered on the celestial equator in the Southern
    Galactic Cap and running from 300$\degree$ to 60$\degree$ in
    R.A. \citep[Sec. 3.2, ][]{Abazajian09}. Co-addition of the best of
    this data means that optical imaging on Stripe 82 can reach roughly 2
    magnitudes fainter than the main survey and as we shall see in
    section~\ref{sec:vhighz}, this deeper imaging was used by
    \citet{Jiang09} to discover new, very high, $z\sim6$, redshift
    quasars. We do not require this deeper imaging for the preparation of
    our SDSS-ULAS matched catalog. However, due to the fact that there is
    almost complete coverage in the ULAS, as well as many other
    multi-wavelength surveys, along with the regular rectangular geometry
    (that will simplify the number counts and luminosity function
    calculations in future work), we utilize heavily the data from this
    area.
    
    \begin{figure}
      \includegraphics[height=8.0cm,width=8.0cm]
      {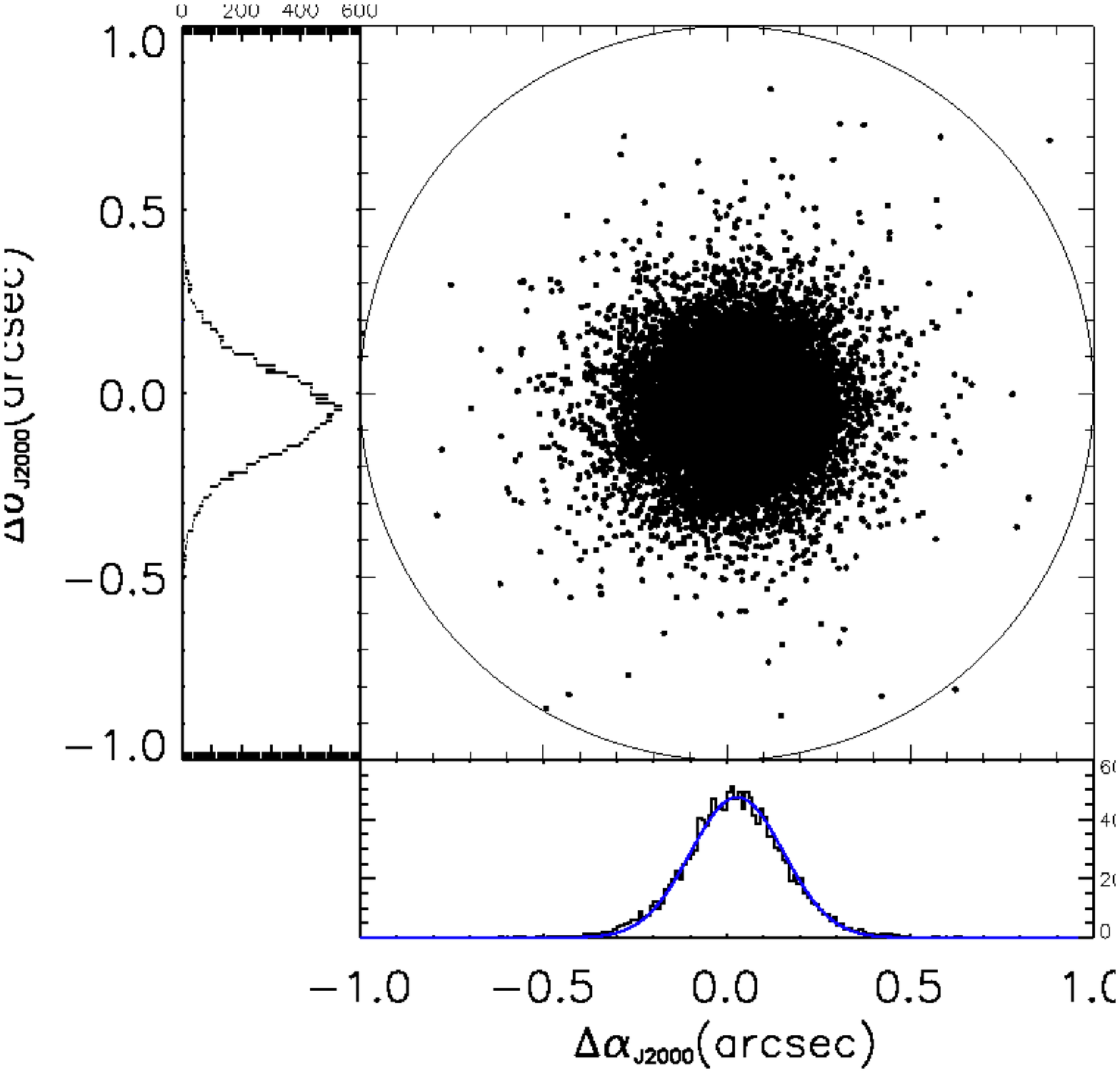}
      \centering
      \caption[Bullseye Plot.]
      {The differences in Right Ascension and Declination
        recorded by SDSS against that recorded by the ULAS for
        all 17,835 matched objects in Stripe 82.
        A circle of radius 1$''$ is shown.}
      \label{fig:radecdif}
    \end{figure}

    \subsection{Matching Procedure}
    Our matching procedure is conceptually straight-forward, though is
    made relatively cumbersome by the shear number of objects to
    process. Before the million strong catalog (DR6pQ) was uploaded, tests
    were perform using the ``CrossID'' form. The positions of the
    1,015,082 objects from the DR6pQ catalog were uploaded to the WFCAM
    Science Archive. To count as a match, the returned NIR object must be
    the {\it closest} object in the ULAS to the SDSS coordinates, selected
    from within a circle of radius 1$''$ (though we note a more stringent
    0.5$"'$ matching radius returns essentially identical results).  As
    long as at least 1 NIR band contained a non-default, i.e. not
    -9.9999$\times 10^{8}$ value, the object was considered a good match.
    Table \ref{tab:totcounts} shows the total number of objects in the R09
    catalog and the number of those objects that are matched to ULAS and
    DXS separately, both for the full DR6pQ coverage and specifically
    Stripe 82.
    
    \begin{table}
      \begin{center}
        \caption{Overall Catalog Numbers.}
        \label{tab:totcounts}
        \begin {tabular}{l|rr} 
          \hline 
          \hline 
          Catalogs & Full DR6 &  Stripe 82 \\ 
          \hline 
          Ri09                                                               & 1,015,082 & 36,625 \\ 
          DR6pQ-ULAS Matches (in $\ge$1 NIR band) &   {\bf 130,827} & 17,835 \\ 
          \hline
          \hline
        \end{tabular}
      \end{center}
    \end{table}
    
    We also perform matching tests where we offset the right ascension
    and declination of Stripe 82 DR6pQ catalog objects and then upload them
    to WSA to observe the probability of spurious matches.  Table
    \ref{tab:secmatch} shows the results of our matching tests.  Once the
    offsets are increased to more than 1'' the amount of spurious
    ``matches'' drops off precipitously.
    
    \begin{table}
      \begin{center}
	\caption{Spurious match counts.  
          Results of secondary matching tests where we offset the right ascension and 
          declination of Stripe 82 DR6pQ catalog objects and then upload them to WSA 
          to observe the probability of spurious matches.}
        \label{tab:secmatch}
        \begin {tabular}{lrr} 
          \hline 
          \hline 
          Offset   &  R.A. offset ``matches'' & Decl. offset ``matches'' \\ 
          \hline 
          1''&  7253&  10965 \\
          2'' &  56 &      40\\
          5'' &  75  &    72\\
          10''  & 84 &     93\\
          60''  & 76 &    78\\
          \hline
          \hline
        \end{tabular}
      \end{center}
    \end{table}        
    
    Figure~\ref{fig:radecdif} displays the separation in arcseconds
    between our matched objects in the SDSS and UKIDSS. The histograms,
    with a bin size of 0.01$''$ running along the sides of the plot,
    represent the distribution of the Right Ascension and Declination
    separations. We find the standard deviation of these histograms to be
    $\delta_{\rm R.A.} = 0.1370''$ and $\delta_{\rm Decl.}= 0.1314''$.
    These values compare very well to those determined in the UKIDSS Early
    Data Release paper \citep[EDR; ][]{Dye06} and by \citet{Chiu07}. Also
    similar to \citet{Chiu07}, we find an absolute offset between the SDSS
    and ULAS of $|{\rm R.A._{offset}}| = 0.025''$, and $|{\rm
      Decl._{offset}}| = 0.040''$. We suggest the origin of this offset to
    be due to the matching of {\it catalogs}, rather than image level
    data, and is not due to the astrometric calibration of either of the
    surveys, nor the size of CCD pixels (the SDSS camera pixel size is 24
    $\mu$m/0$''$.396 on the sky; WFCAM pixels are 18$\mu$m/0$''$.4 on the
    sky). However, we also suggest this issue warrants further
    investigation.

\begin{table*}
  \begin{center}
    \caption{The first ten objects in our NIR-matched quasar catalog.
      The catalog will be published in its entirety in the electronic 
      edition of the Astrophysical Journal. The first
      10 rows, of 130,827 total, are shown here.}
    \label{tab:first_ten}
    \begin {tabular}{rrccrrrrr} 
      \hline 
      \hline 
      No.  & Name & SDSS ID & ULAS ID & R.A.$_{\rm SDSS}$ & Decl.$_{\rm SDSS}$ &  $\Delta_{\rm R.A.}$ &  $\Delta_{\rm Decl.} $  & Diff\\
      \hline 
      1 & 000001.38-010852.2 & 588015507658768592 & 433825841152 & 0.0057816 & -1.1478427 &  -0.180  &   0.001 & 0.180 \\ 
      2 & 000001.93-001427.5 & 588015508732510344 & 433822400512 & 0.0080576 & -0.2409745 &  -0.130  & -0.137 & 0.189 \\ 
      3 & 000006.42+005206.3 & 588015510343123177 & 433816469504 & 0.0267757 &  0.8684207 &     0.031  & -0.316  & 0.318 \\
      4 & 000006.53+003055.2 & 588015509806252150 & 433817550848 & 0.0272281 &  0.5153489 &     0.118  &  0.164 & 0.202 \\
      5 & 000007.58+002943.3 & 588015509806252166 & 433817550848 & 0.0316036 &  0.4953744 &   -0.039  &  0.127 & 0.133 \\
      6 & 000008.13+001634.6 & 588015507658768713 & 433819123712 & 0.0338984 &  0.2763040 &   -0.077  & 0.108 &  0.132\\
      7 & 000009.31-010703.1 & 588015507658768768 & 433825644544 & 0.0387928 & -1.1175526 &  -0.050  &  0.105 & 0.117 \\
      8 & 000011.96+000225.3 & 588015509269381139 & 433820860416 & 0.0498422 &  0.0403718 &  -0.064   & -0.078 & 0.101 \\
      9 & 000012.25-003220.5 & 587731185667080338 & 433823776768 & 0.0510761 & -0.5390492 &    0.033  & -0.316 & 0.317 \\
    10 & 000012.27-010405.5 & 588015507658768732 & 433825841152 & 0.0511651 & -1.0682041 &  -0.062  &  0.193 & 0.203 \\
      \hline
      \hline
    \end{tabular}
  \end{center}
\end{table*}    

\begin{table*}      
  \centering 
  \caption{Format of our Matched catalog. 
    With field names and descriptions.}
  \label{tab:cat} 
  \begin{tabular}{@{}rlll@{}} 
    \hline 
    \hline 
    Column & Field&Format&Description\\ 
    \hline
    1 & Catalog number	& I5 & Internal catalog object number\\ 
    2 & Name             	& A18 & IAU format object name; SDSS Jhhmmss.ss$\pm$ddmmss.s \\
    3 & SDSS - ID       	& I18 & SDSS Database ID \\
    4 & UKIDSS - ID   	& I12 & ULAS MergedSource Database ID\\ 
    5 & R.A. SDSS        	& F12.7 & R.A.  J2000 (deg)\\ 
    6 & Decl. SDSS      	& F12.7 & Decl. J2000 (deg)\\
    7 & $\Delta$ R.A. 	& F12.7 & Difference Between R.A.  from SDSS and UKIDSS (arcsecs)\\ 
    8 & $\Delta$ Decl.       & F12.7 & Difference Between Decl. from SDSS and UKIDSS (arcsecs)\\
    9 & Diff                	& F12.7 & Difference between UKIDSS Object and SDSS Object designated as a Match (arcsecs)\\ 
    10 & $u$	        	& F12.7 & $u$ SDSS PSF magnitude in AB\\
    11 & $u$ err	       	& F12.7 & $u$ Error SDSS PSF magnitude in AB\\
    12 & $g$	        	& F12.7 & $g$ SDSS PSF magnitude in AB\\
    13 & $g$ err	       	& F12.7 & $g$ Error SDSS PSF magnitude in AB\\
    14 & $r$	        	& F12.7 & $r$ SDSS PSF magnitude in AB\\
    15 & $r$ err	        	& F12.7 & $r$ Error SDSS PSF magnitude in AB\\
    16 & $i$	        	& F12.7 & $i$ SDSS PSF magnitude in AB\\
    17 & $i$ err	        	& F12.7 & $i$ Error SDSS PSF magnitude in AB\\
    18 & $z$       		& F12.7 & $z$ SDSS PSF magnitude in AB\\
    19 & $z$ err   		& F12.7 & $z$ Error SDSS PSF magnitude in AB\\
    20 & $Y$	        	& F12.7 & $Y$ ULAS PSF magnitude in Vega\\
    21 & $Y$ err	       	& F12.7 & $Y$ Error ULAS PSF magnitude in Vega\\
    22 & $J$	        	& F12.7 & $J$ ULAS PSF magnitude in Vega\\
    23 & $J$ err	        	& F12.7 & $J$ Error ULAS PSF magnitude in Vega\\
    24 & $H$	        	& F12.7 & $H$ ULAS PSF magnitude in Vega\\
    25 & $H$ err         	& F12.7 & $H$ Error ULAS PSF magnitude in Vega\\
    26 & $K$	        	& F12.7 & $K$ ULAS PSF magnitude in Vega\\
    27 & $K$ err	        & F12.7 & $K$ Error ULAS PSF magnitude in Vega\\
    28 & depth$\_Y$	      	& F12.7 & nominal 5$-\sigma$ depth in the $Y$-band \\
    29 & depth$\_J$	      	& F12.7 & nominal 5$-\sigma$ depth in the $J$-band \\
    30 & depth$\_H$    	& F12.7 & nominal 5$-\sigma$ depth in the $H$-band \\
    31 & depth$\_K$	      	& F12.7 & nominal 5$-\sigma$ depth in the $K$-band \\
    32 &zPhot	        	& F12.7 & Photometric Redshift \\
    33 &zPhotLow             & F12.7 & Lower limit of the photometric redshift range \\
    34 &zPhotHi                & F12.7 & Upper limit of the photometric redshift range \\
    35 &zSpec	        	& F12.7 & Spectroscopic Redshift (when known, $-1$ otherwise) \\
    36 &zphotprob     	& F12.7 & Probability Photometric Redshift is accurate\\
    37 &E(B-V)	         	& F12.7 & Galactic Extinction\\ 
    \hline 
    \hline 
  \end{tabular} 
\end{table*}

\section{The Matched Catalog} 
Our matched catalog between the SDSS DR6pQ and the ULAS DR3 (hereafter
simply referred to as ``the matched catalog'') comprises a total of
\hbox{130,827} objects matched between the R09 catalog
and the ULAS.  The first ten objects are given in
Table~\ref{tab:first_ten}, with the column titles, meanings and format
given in the following text and in Table \ref{tab:cat}.  The full
version is found in the electronic version in machine readable form.

{\bf Columns 1 through 4} The first four columns designate the
internal catalog number; the formal name of the quasar as reported
from the SDSS, with the form at SDSS Jhhmmss.s$\pm$ddmmss.ss; and the
ObjID numbers specific to the SDSS and UKIDSS, specifically from the
{\tt PhotoObjAll} and {\tt lasSource} database tables respectively.

{\bf Columns 5 and 6} give the J2000 right ascension and
declination from SDSS in degrees.

{\bf Columns 7 through 9} shows the relative positional measurement
accuracy between SDSS and UKIDSS. $\Delta$R.A. and
$\Delta$Decl. values are calculated by subtracting positional
measurement as recorded by SDSS by the positional measurement as
recorded by UKIDSS. Distance values are not calculated explicitly,
rather they are values reported directly from the WFCAM database.  The
``distance'', i.e. difference, $|d|$, is simply given by:
\[
   d = \sqrt{( {\rm R.A.}_{\rm SDSS}   - {\rm R.A.}_{\rm UKIDSS}) ^2 +
                  ( {\rm Decl.}_{\rm SDSS}  - {\rm Decl.}_{\rm UKIDSS})^2}
\]		
Distances are the only angular quantities expressed in arcseconds
rather than degrees.
    
{\bf Columns 10 through 19} give the PSF magnitude values in each of
the 5 SDSS optical bands, and their associated error. Errors are not
re-calculated, but are those previously reported in R09.

\begin{table*}
  \centering
  \caption{Counts for each of the NIR Bands in the catalog,
    along with colors and count in all 4 NIR Bands in bold.  The
    italic numbers in parentheses are the totals for Stripe 82 only.}
  \label{tab:counts}
  \begin{tabular}{lrrrrrrr}
    \hline 
    \hline 
    Band & $Y$ & $J$ & $H$ & $K$ & $J+H$ & $J+K$ & $J+H+K$ \\  \hline	
    $Y+$ & 80,544 (\it{13,681}) & 59,085 (\it{9,384}) & 54,965 (\it{8,690}) & 53,656 (\it{8,442}) & 47,787 (\it{7,258}) & 46,338 (\it{6,938}) & \bf{42,133 (\it{6,007})} \\ 
    $J+$ & $--$ & 89,962 (\it{11,141}) & 51,701 (\it{8,236}) & 50,264 (\it{7,993}) & $--$ & $--$ & $--$ \\ 
    $H+$ & $--$ & $--$ & 72,347 (\it{11,191}) & 61,015 (\it{8,892}) & $--$ & $--$ & $--$ \\ 
    $K+$ & $--$ & $--$ & $--$ & 74,351 (\it{11,546}) & $--$ & $--$ & $--$ \\
    \hline
    \hline
  \end{tabular}
\end{table*}

{\bf Columns 20 through 27} provide NIR magnitudes, {\it
j\_1AperMag3}, {\it yAperMag3}, {\it hAperMag3} and {\it kAperMag3},
and the associated error as given in the ULAS database.  The
AperMag3 magnitudes are the aperture corrected magnitudes measured by
UKIDSS, which records all the flux in a 2.0$''$ diameter and that for
typical seeing conditions, provide the most accurate estimate of the
total magnitude \citep{Dye06}. The {\it j\_1AperMag3} magnitude is
reported, since the ULAS ultimately aims to have {\it J}-band imaging
in two epochs.  For quasars lacking coverage in one or more bands, the
default reported magnitude and error are \hbox{-9.9999$\times
10^{8}$}.  Table \ref{tab:counts} shows the breakdown of matches based
on each NIR band and combination of bins. Objects numbers for Stripe
82 only are given in parentheses.

{\bf Columns 28, 29, 30 and 31} give the nominal 5-$\sigma$ depth,
$m_{d}$, at the given coordinates for a R09 quasar catalog member that
falls within the ULAS DR3 footprint (in any one of the four NIR
bands).  Following equation (5) from \citet{Dye06}, $m_{d}$ is the
5$\sigma$ detection limit for a point source, defined as five times
the standard deviation of the counts in an aperture, corrected for
flux outside the aperture:
\[
m_{{\rm d}} =m_{{\rm 0}} - 2.5 \log_{10} ( 5\sigma (1.2 N_{p})^{1/2}/ t_{\exp} ) - m_{{\rm ap}}
\]
where $m_{0}$ is the photometric zero-point (the {\tt PhotZPCat}
attribute in the WSA); $\sigma_{\rm sky}$ is the measure of the
standard deviation of counts in the sky ({\tt skyNoise}); $N_{p}$ is
the number of pixels in the aperture, the factor of 1.2 accounts for
the covariance between pixels and $t_{\rm exp}$ is the Exposure Time,
({\tt expTime}). $m_{{\rm ap}}$ is the aperture correction term, {\tt
AperCor3}, which gives the correction needed for a 2.0$"$ aperture
diameter (i.e. the respective quantity for the quoted aperMag3
values).  These values can all be found in the {\tt Multiframe} and
{\tt MultiframeDetector} table of the WSA. Section 10, and in specific
section 10.2.2, example 3, in \citet{Dye06} has further details here.
  
{\bf Columns 32 through 34} represent the photometric
redshift\footnote{On occasion, a photometric redshift of -1 is
returned in the catalog; this is simply an error saying that the
photo-$z$ code failed for some, usually tractable reason
(G.T. Richards, priv. comm.). This issue only affects 3,574 objects
out of the total 1,015,082 from R09 quasar catalog and
only 242 quasars on Stripe 82.  We do not attempt to deal with this
issue, leaving the resolution for the next version of the photometric
catalog.}  from R09, with columns 29 and 30 giving the
lower and upper limits of the photometric redshift range
respectively. These values should be taken with the {\tt zphotprob}
quantity, described below.  Note that these upper and lower redshift
ranges a nearly, but ultimately not, symmetric about the reported
photometric redshift, and therefore should not be treated as the
formal error on the photometric redshift.

{\bf Column 35} gives the spectroscopic redshifts are from the 
R09 catalog, and are ultimately based on matches to the SDSS DR5 quasar
catalog \citep{Schneider07}, the 2QZ quasar catalog \citep{Croom04},
the 2dF-SDSS LRG and QSO Survey (2SLAQ) quasar catalog
\citep{Croom09a}, and the SDSS DR6 spectroscopic database
\citep{Adelman-McCarthy08}.  By design the R09 catalog
does not have spectroscopic information for all of its objects. We
find that out of the \hbox{130,827} objects matched from R09 and
ULAS, \hbox{20,740} have spectroscopic redshifts as reported in
R09, \hbox{6,623} of which are on Stripe 82. The value $-1$ is given when
there is no corresponding spectroscopic redshift.

{\bf Columns 36} represent the probability of the reported photometric
redshift being in the given redshift range, with R09 (their section
4.6 and Figure~15), and section 4 in our investigations, giving
further details on the {\tt zphotprob} quantity.

{\bf Column 37} represents the estimated Galactic reddening at the
given position from \citet{Schlegel98}.

The smaller UKIDSS DXS DR3 matched catalog, with \hbox{1,070} matched
objects, has exactly the same format as given here for the SDSS-ULAS
matched catalog given here, and again will be available with the
electronic version. The only notable difference being that since the
DXS does not cover the $Y$-band, all those magnitudes are reported as
0.00.

\section{Global Catalog Properties}
    
    \subsection{Coverage}
    Only around 10-14\% of the R09 objects match to the ULAS DR3, with
    the primary reason for this low percentage being the difference in
    coverage between the SDSS DR6 and the ULAS DR3
    (Figure~\ref{fig:fullcov}) gives the ULAS DR3 coverage). To
    investigate how many R09 objects actually lie with-in the ULAS DR3
    footprint, we take the R09 catalog and the ``multiframe'' information
    connected to UKIRT via the WSA. A multiframe is essentially the
    footprint of the four detectors of an individual WFCAM exposure
    \citep[see e.g. section 10, ][]{Dye06}. In total there were
    \hbox{172,186} R09, optically detected objects that fall within the
    ULAS DR3 footprint, in the union of the 4 bands.  For individual
    bands, there are \hbox{109,959}, \hbox{141,304}, \hbox{132,788} and
    \hbox{135,257} matches in the {\it Y}, {\it J},{\it H} and {\it K}
    bands respectively, with \hbox{97,541} R09 objects were in the
    intersection of the {\it Y}, {\it J}, {\it H} and {\it K } coverage
    footprint. Thus, and using Table~\ref{tab:counts}, we find that 73\%,
    64\%, 54\% and 55\% of the SDSS R09 objects are detected in the DR3
    footprint, in the {\it Y}, {\it J}, {\it H} and {\it K} bands
    respectively. 43\% of the objects have detections in the all four NIR
    filters.
    
    The surface density is $\approx$122 deg$^{-2}$ for the R09
    DR6pQ. This compares to the surface density for objects with $K\leq
    18.27$ (see discussions below regarding this limiting magnitude) of
    $\approx$80 deg$^{-2}$. These values are generally in line with those
    given in \citet{Smail08} who derive a surface density of QSOs with
    $K\leq 20$ of between 85--150 deg$^{-2}$.

    \begin{figure}
      \includegraphics[height=8.0cm,width=8.0cm] 
      {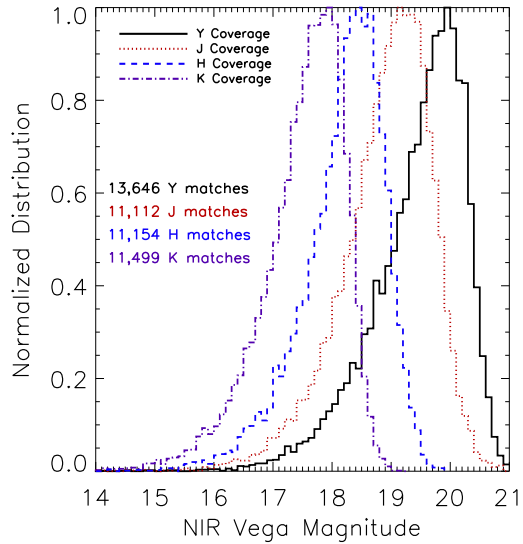} 
      \centering 
      \caption[Distributions of each of the four NIR bands from Stripe 82
      ULAS matches in {\it Vega} magnitudes.]
      {Distributions of each of the four NIR bands from Stripe 82
        ULAS matches in {\it Vega} magnitudes, $Y$ ({\it solid black}), $J$,
        ({\it dotted red}), $H$ ({\it dashed blue}) and $K$ ({\it dot-dashed
          purple}), for the matched objects, normalized at the peak to unity.}
      \label{fig:nirhist}
    \end{figure}
    \begin{figure} 
      \includegraphics[height=8.0cm,width=8.0cm] 
      {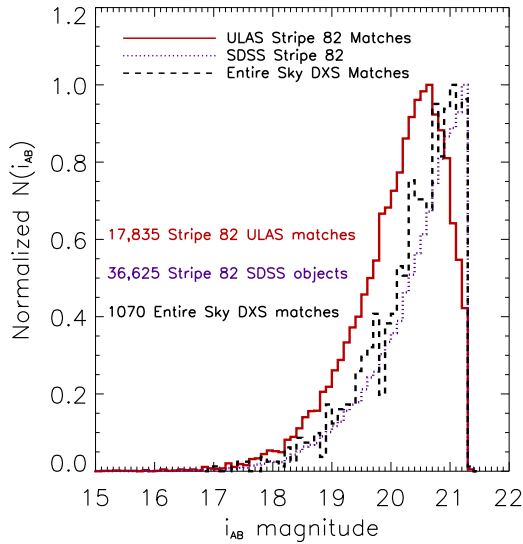}  
      \centering 
      \caption[Histogram showing population by $i$-band PSF magnitude.] 
      {Histogram showing population by $i$-band PSF magnitude.  Each
        line was normalized to a peak value of 1.0 and a bin size of 0.1. The
        purple dotted line shows the histogram based on objects from R09 in
        Stripe 82, the solid red line displays the histogram of only ULAS matched
        objects and the black dashed line represents the histogram of DXS
        matches over the entire sky.} 
      \label{fig:ihist} 
    \end{figure} 
    \subsection{Magnitude Distributions}
    Figure~\ref{fig:nirhist} shows the normalized magnitude
    distributions for the matched objects in each of the four NIR bands,
    showing at which magnitude each particular band is limited by.  As
    expected, each band peaks at a different value with shorter wavelength
    bands peaking at fainter magnitudes.
    
    For Figure~\ref{fig:ihist}, the solid (red) histogram represents
    the objects which are matched in any of the 4 NIR bands, while the
    dotted (purple) histogram represents all the optically detected
    objects from the DR6pQ on Stripe 82 only.  The dashed (black) line are
    the matches to the deeper DXS, the limiting depth of which is $K=20.8$
    \citep{Lawrence07, Stott_thesis, Swinbank07}.  Due to the smaller
    amount of available data for UKIDSS DXS, the matches characterized by
    the dashed line is less smooth than the other two histograms. The
    sharp cut-off at $i$=21.3 displays the $i$-band limit in the original
    SDSS DR6 photometric redshift catalog.
    
    The $i$-band magnitude distribution of the matched catalog
    strongly suggests that the ULAS is not complete at the optically faint
    end of DR6pQ. The suggested completeness limit for the matched catalog
    is thus $i\approx 20.6$. We also caution that our matched catalog is a
    heterogenous sample of objects, and {\it not a complete statistical
      sample} as is. This will have particular ramifications when the
    calculation of the NIR Quasar Luminosity Function (QLF) comes to be
    calculated in our companion paper (Peth, Ross et al., in prep.).  The
    significant shift between the ULAS matched histograms and the other
    two (optical-Stripe 82 and DXS matched) distributions most likely
    implies brighter optical objects are preferentially matched to ULAS
    NIR detections. It is reassuring to see that the DXS and optical-only
    histograms seem to match very well in shape and limiting magnitude,
    and the matched DXS catalog is provided as a test-bed for future
    investigations into the fainter, $i\gtrsim 20.6$ end of the $K$-band
    quasar population.

   \begin{figure} [h]
      \includegraphics[height=8.0cm,width=8.0cm] 
      {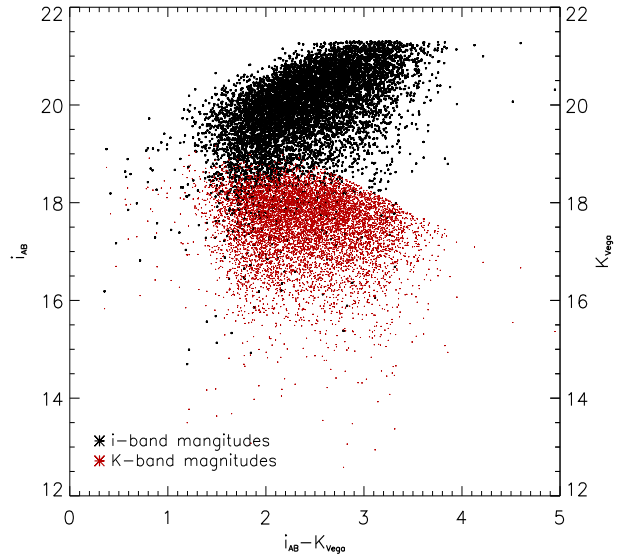}  
      \centering 
      \caption[] 
      {Color-magnitude histograms for both the $i_{\rm AB}$ and the $K_{\rm Vega}$ bands, 
        for $i-K$ color. Note the lack of points with faint ($i>20$) blue, $(i-K)<2.5$,
      colors and the effect of the limiting $K$-band magnitude for redder, 
      $(i-K)\gtrsim2.7$, objects.}
      \label{fig:ik}
    \end{figure}
    \begin{figure*}
      \includegraphics[height=15.0cm,width=15.0cm] 
      {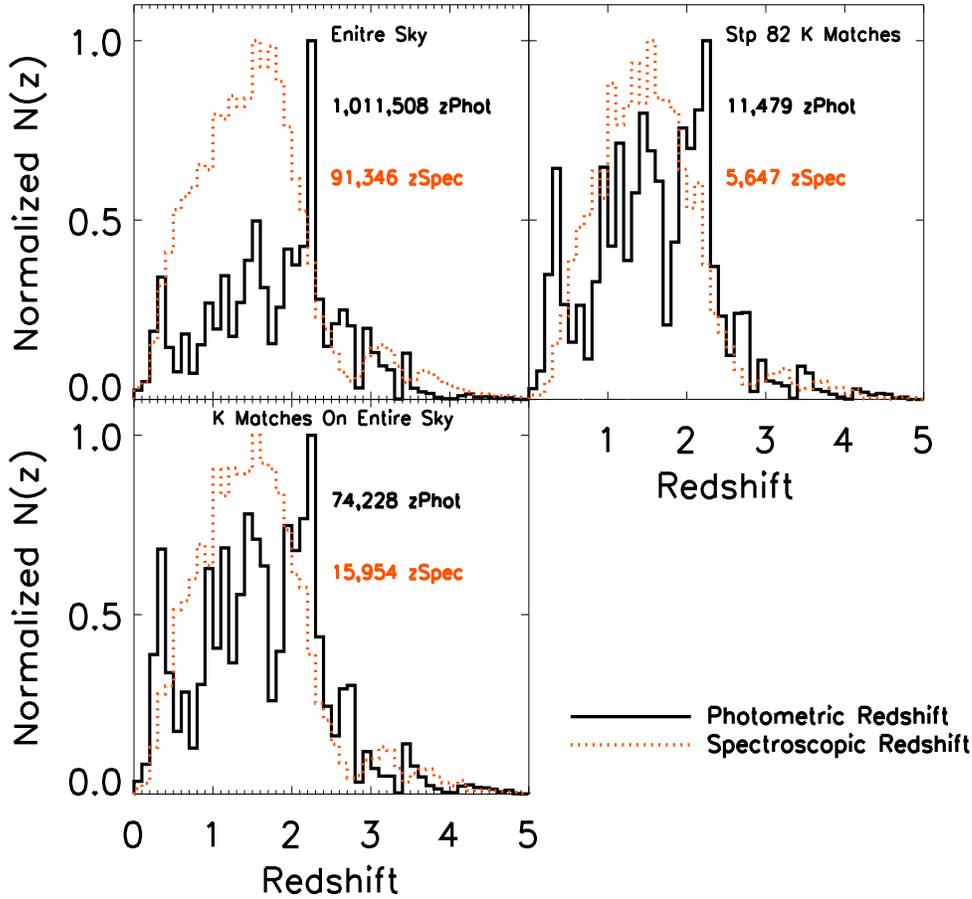}
      \centering 
      \caption[Distributions of Photometric and Spectroscopic Redshifts]
      {Distributions of Photometric and Spectroscopic Redshifts, for
        the full DR6 coverage ({\it upper left}); Stripe 82 only ({\it upper
          right}); $K$-band matches over the DR6 ({\it lower left}).  The black
        line represents Photometric redshift distributions and the orange line
        represents spectroscopic redshift distributions.  The subsample of
        quasars from Stripe 82 (the upper-right plot) posses nearly identical
        redshift distributions to those over the entire sky.  However, objects
        with $K$-band coverage have photometric redshifts that are
        preferentially in the low redshift bins.  The difference in numbers
        given here and in Table~\ref{tab:counts} are due to the sample number
        of objects which have a photometric redshift of {\tt zPhot}$=-1$.}
      \label{fig:redhist}
    \end{figure*}
   
    Figure~\ref{fig:ik} gives the color-magnitude distributions for
    both the $i$ (black points) and $K$ (red points) bands. Again the
    sharp cut-off at $i=21.3$ is seen. The $K$-band limit is seen to be
    down to $\approx$18.4, potentially a little deeper, consistent with
    Figure~\ref{fig:nirhist}. This is slightly deeper than the preliminary
    depths reported in \citet{Lawrence07} but in good agreement with the
    updated (UKIDSS DR2) values from \citet{Warren07DR2} and indeed is
    also consistent with the measurements from the ULAS team, who give a
    5$\sigma$ point source limiting magnitude of 18.27 (R.G. McMahon,
    priv. comm.). We suggest these faintest $K$-band objects are actually
    5$\sigma$ detections in a non-$K$ NIR band, but then also have some
    flux in the aperture used to calculate the {\it kAper3Mag}, just
    possibly not to as strong a limit as the quoted 5$\sigma$ for point
    sources. Also apparent are the clear selection affects associated with
    the matched catalog, with few faint $i$-band objects at
    $(i-K)\lesssim2$ and very few objects fainter than $K=18$ and redder
    than $(i-K)\gtrsim3$.  Again, the location of the matched catalog
    objects in e.g. $K$ vs. $(i-K)$ color-magnitude space, will become
    critical for statistical calculations using these data.
    
    As a check, we visually examine (via the SDSS CAS Image List Tool
    and the UKIDSS WSA GetImage form) the very reddest objects in our
    catalog. These very red objects are defined as having $K$-band
    detections and colors of $(r-K)\geq 5.0$ or $(i-K)\geq 4.5$. There
    criteria return 17 and 10 objects respectively. Upon visual inspection
    of these 17 (10) sources, 6 (4) turned out to be either associated
    with, or contaminated by a foreground extended source. However, 13 out
    of the 17 $(r-K)$ selected objects had suspiciously high, $z=4.5-5.6$,
    redshifts, or {\it zphotprob} $\leq0.8$ values, indicating the
    deceptive nature of these objects. Of the 10 $(i-K)$ selected sample,
    the 4 sources associated with the foreground galaxies, also had {\it
      zphotprob} $\leq0.8$ values, while the remaining 6 objects appeared as
    genuine point sources. These particular very red $(i-K)$ objects
    deserve future follow-up investigations and we shall discuss in more
    detail in the next section the utility of the {\it zphotprob} flag.
    
    
    \begin{figure*}
      \includegraphics[height=15.0cm,width=17.0cm] 
      {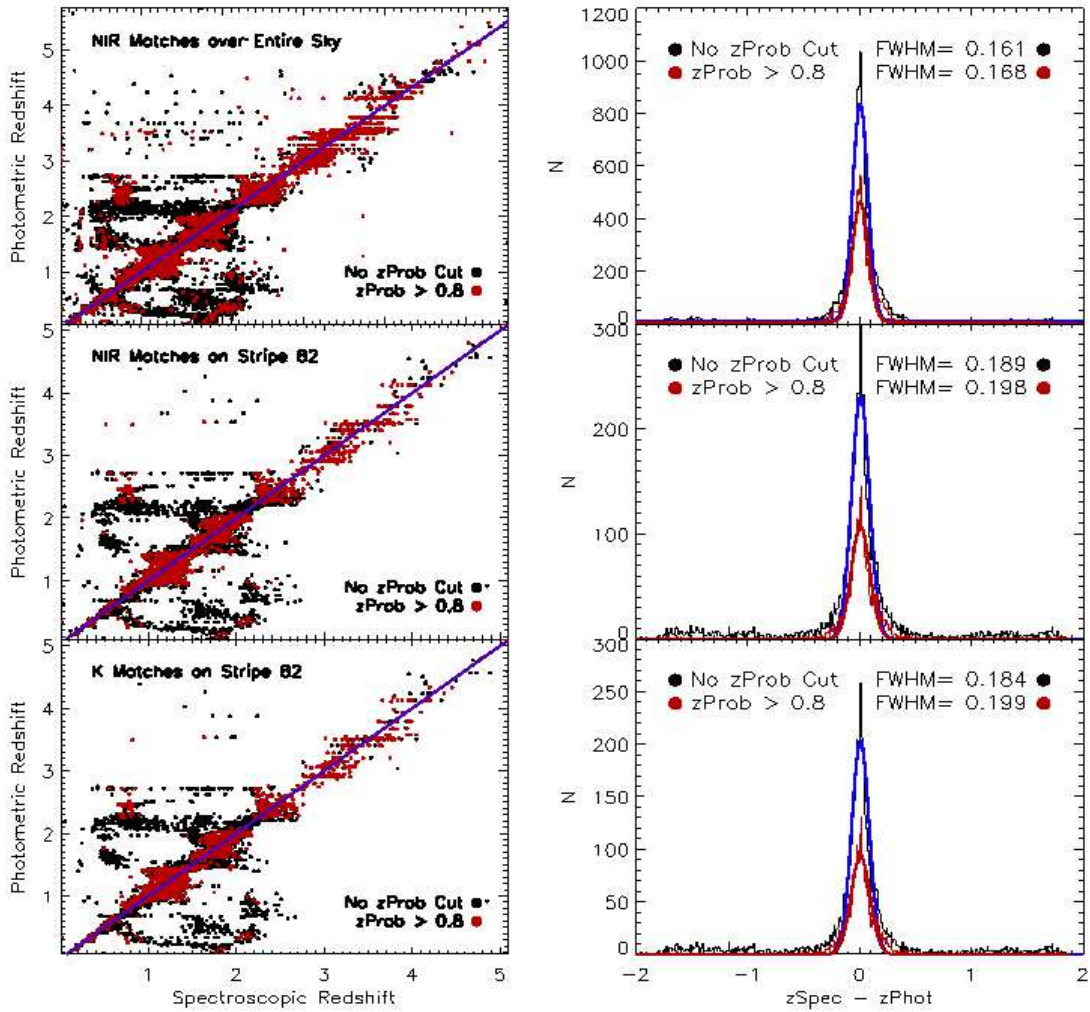} 								
      \centering 
      \caption[] 
      {{\it Left panel}: 
        The spectroscopic redshifts plotted against photometric
        redshift for all matches with spectroscopic redshifts.  Black points
        have no {\tt zphotprob} cuts while red points have {\tt zphotprob}$>0.8$.  
        The 1:1 relation is plotted with a purple line.
        {\it Right panel}:
       Histograms of difference between zspec and zphot for (1) NIR
       matches over the entire sky ({\it upper right}); (2) NIR matches on
       Stripe 82 ({\it middle right}); (3) $K$- matched objects on Stripe 82
       ({\it lower right}).  A small full width half maximum says that the
       majority of photometric and spectroscopic redshifts are in agreement.} 
      \label{fig:zspecphot} 
    \end{figure*}
    
    \subsection{Redshift Distributions}
    Figure~\ref{fig:redhist} shows the redshift distribution, $N(z)$,
    for quasars, in redshift bins of $\delta z =0.1$. Three scenarios are
    presented; the top left panel is for the full, optical-only, SDSS DR6
    photometric redshift catalog; the lower left panel is from the matched
    catalog with $K$-band detections across the full coverage, and top
    right is from the matched catalog with $K$-band detections from Stripe
    82.  In each panel, the redshift distribution of the photometric
    redshifts is given by the solid (black) line, while those objects with
    spectroscopic redshifts are shown by the dashed (orange) histogram.
    
    For the optical-only distributions the redshift distributions are
    potentially well matched, apart from the excess of photometric
    $z\sim2.3$ objects which distort the normalization.  
    
    The redshift histograms for those objects with $K$-band detections
    from the matched catalogs, for both the full coverage and Stripe 82,
    are almost identical. Interestingly, the spike in the number of
    objects at $z\sim2.3$ seems to be dramatically reduced (although does
    not disappear completely) for the {\it K}-band matches.  As we shall
    see in $\S$ 4.6, when we examine the stellar contamination of our
    sample using the {\it giK} color-color space, we find that {\it (i)}
    stellar contamination is potentially higher in objects with
    photometric redshifts $z\sim2.3$ and {\it (ii)} due to the respective
    shape of typical quasar and (e.g. M-type) stellar SEDs
    \citep[e.g. Figure 1, ][]{Maddox08}, quasars will be preferentially
    selected over stars, if measured in the {\it K}-band. Thus, we suggest
    the $z\sim2.3$ spike is because these mid-redshift objects have a
    lower ``zphotprob'' value than the general photometric quasar
    sample. As explained later in $\S$ 4.6 we see that the stellar locus
    separates nicely for $z \lesssim$ 3.2 but not for higher redshifts.
    
    Once the $K$-band matches are made, the photometric and
    spectroscopic histograms now seem to be in reasonable agreement,
    though there are possibly decrements of photometric objects at
    $z\sim0.7$ and more severely at $z\sim1.8$. We note that this is not a
    new feature, having been seen in R09 (their Figure 14), and as such are
    motivated to continue investigations for the photometric and
    spectroscopic redshift distributions from our matched catalog.
    
    \begin{figure*}
      \includegraphics[height=15.0cm,width=15.0cm] 
      {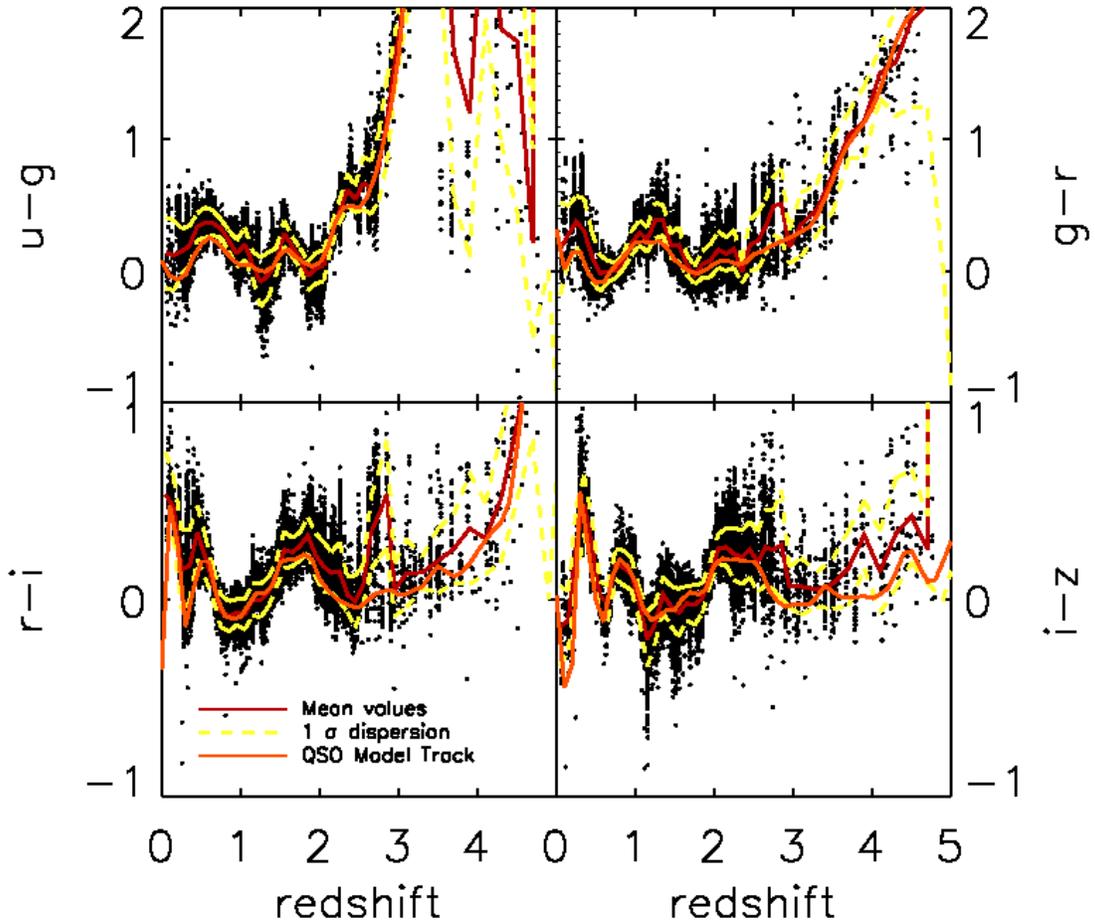} 
      \centering 
      \caption[Color versus redshift plots from Stripe 82 matches of $u-g$,
        $g-r$, $r-i$ and $i-z$.] 
      {Color versus redshift plots from Stripe 82 matches of $u-g$,
        $g-r$, $r-i$ and $i-z$.  Mean magnitude ($red$ $line$) in a binsize of
        0.1 mag, 1$\sigma$ dispersion ($striped$ $yellow$ $line$) both above
        and below the mean values, and Model Quasar colors ($orange$ $line$)
        as determined in \citep{Hewett06} are each plotted in terms of
        redshift.} 
      \label{fig:4col} 
    \end{figure*}
    Figure~\ref{fig:zspecphot} shows the relationship between the
    photometric redshifts and the spectroscopic redshifts for the objects
    in our matched catalog. From top to bottom, the left panel of figure
    \ref{fig:zspecphot} is; the entire matched catalogue with
    spectroscopic coverage; all NIR matches on Stripe 82 and all $K$-band
    matches on Stripe 82. A purple line shows the 1:1 relation between
    spectroscopic and photometric redshifts. There is considerable
    structure in the photometric-spectroscopic quasar redshift relation,
    which again has been seen previously, e.g. \citet{Mountrichas07}, R09.
    Degeneracies between $z\sim0.8$ and $z\sim2.4$, as well as the width
    of the distributions around $z\sim1-1.5$ can be understood by the
    appearance, and potential confusion of, various strong emission lines
    in the various (SDSS) filters \citep[see e.g. ][for detailed
    descriptions]{Fan99, Richards02, Richards04}.  There are a couple of
    notable differences between the left hand panels of
    Figure~\ref{fig:zspecphot} that should be briefly highlighted.  First,
    note that a number of objects with high, $z>3$, photometric redshifts
    and no {\it zphotprob} cut, disappear when there is a match to the NIR
    data.  Upon checking, we suggest this is caused by these objects
    generally being lower luminosity AGN than ``standard'' quasars, and
    thus thought to be at higher-$z$ than was spectroscopically found, and
    in general just too faint for the ULAS to detect. Also, though, this
    is good motivation for the utilization of the {\it zphotprob} flag.  A
    second feature is the continued structure in the NIR matched
    plots. This, coupled with the general lack of structure at $z>3$,
    suggests that as far as photometric redshift estimation for quasars
    (and not e.g. differentiation between high-$z$ quasars and cool
    stars), then the directly matched NIR photometry, i.e. ``Is there
    simply NIR photometry?'', actually does not provide a great amount of
    extra information. Of course this does not take into account the
    actual detected NIR fluxes (or upper limits) that would help refine
    photo-$z$ estimates, or the use the NIR to find very high,
    $z\gtrsim6$, redshift objects.

    \begin{table*}
      \centering
      \caption{The number of objects with photometric and spectroscopic that 
        agree to within a 3$\sigma$ difference.  The total number of objects in 
        each bin is displayed in $italics$.}
      \label{tab:agree}
      \begin{tabular}{lrcrcrc}
        \hline 
        \hline 
        & SDSS DR6 & $\%$ & Stripe 82 &  $\%$ & K-matched &  $\%$ \\ \hline
        zprob $>$ 0.0 & 15527 (\it{20605}) & 75 & 4933  (\it{6589}) & 75 & 4218  (\it{5644}) & 75 \\
        zprob $>$ 0.8 & 8981  (\it{10188}) & 88 & 2403  (\it{2620}) & 92 & 2114  (\it{2292}) & 92 \\
        \hline
        \hline
      \end{tabular}
    \end{table*}

    The right panel of figure \ref{fig:zspecphot} shows the histogram
    and gaussian fits with calculated full width half maxima for the
    difference in photometric and spectroscopic redshifts.  We define that
    quasars with photometric and spectroscopic redshifts that have a
    difference greater than 3$\sigma$ to be ``catastrophic failures''.
    All quasars with a difference in photometric and spectroscopic
    redshifts that fall within the 3$\sigma$ are deemed ``non-catastrophic
    failures''.  The full numbers on the agreement between spectroscopic
    and photometric redshifts can be seen in table \ref{tab:agree}.
    
    Red points in the left panel of figure \ref{fig:zspecphot} show
    quasars after a redshift probability, {\tt zphotprob} cut of 0.8, the
    value of 0.80 being inspired by the checks done by R09 (their
    Figure 15).  This cut reduces the amount of catastrophic failure
    redshifts by a considerable margin, although the FWHM of the
    ($z$Spec-$z$phot) difference only changes very marginally. Although
    the {\it zphotprob}$\geq 0.80$ is a ``blunt'' cut, we suggest this as
    a relatively sensible division, but do note there are recent, more
    sophisticated studies, e.g. \citet[][]{Myers09}, that show how one can
    take into account the full photometric redshift probability
    distribution function when performing statistical calculations.

    \subsection{Optical and NIR Colors}
    We now turn to the optical and NIR colors of our matched catalog.
    
    Figure~\ref{fig:4col} shows the optical colors, $(u-g), (g-r),
    (r-i)$ and $(i-z)$, as a function of redshift, for our matched catalog
    on Stripe 82. The mean colors of the data are given by the solid (red)
    line, and were determined in a given bin of width $\delta z=0.1$, with
    the associated 1$\sigma$ standard deviations given by the striped
    (yellow) line. The orange triangular-pointed line represent the
    ``Average'' QSO model colors presented in \citet{Hewett06}.
    \citet{Hewett06} also present ``blue'' and ``red'' model quasar
    spectra, where the power-law slope, $\alpha$, at wavelengths
    $\lambda<$12,000\AA\, is $\alpha=0.0, -0.6$ and $-0.3$ for blue, red
    and average model quasars respectively. We cut the plots at redshift
    $z=3$, since we feel we have poorer statistics at higher redshifts and
    that there are not enough points to render a representative mean
    value. We will examine the colors of individual high-$z$ quasars in
    Section~\ref{sec:highz}.
    
    Previous studies \citep[e.g.][]{Richards01, Jester05, Richards06,
      Croom09a, Hewett10} have provided detailed studies of the features of
    quasar colors in the SDSS color space, and their form as a function of
    redshift. As such, we will not repeated those analyses here. Instead,
    we will make comparisons to the models presented in \citet{Hewett06}
    of which details can be found in \citet{Maddox06}.
    
    Starting with $(u-g)$, there are potentially two redshift ranges,
    $z=0-0.6$ and $z>2.3$ where the mean colors of the matched quasars are
    generally {\it redder} than the model colors. While the differences
    are relatively small, $\sim0.2$ mag at the low redshift end, $\sim0.1$
    mag at the high redshifts, the model tracks are consistently bluer
    than the observed quasars, and at $z<0.6$, barely consistent with the
    1$\sigma$ spread of the quasar $(u-g)$ color. Note also however, that
    care has to be taken in the determination of the $(u-g)$ color at
    $z>2.2$ since this is where the Lyman-$\alpha$ forest first enters the
    optical bands, and the determination of accurate $u$-band photometry
    becomes increasingly problematic.
    
    For $(g-r)$, the model tracks reproduce the mean quasar colors
    well at all redshifts, $z=0-3$. However, there might be a tend for the
    observed quasars to be generally redder at low ($z<0.5$) and high
    ($z>2.3$) redshift, though this can be seen as marginal.
    
    For $(r-i)$, it appears that the model tracks do in fact reproduce
    the mean quasar colors very well, for $z<2$, and especially for
    $z=0.7-2.0$.  The disagreement between model and observed mean at
    $z\approx0.5$ seems to be caused by the fact that there are a
    noticeable number of (red and blue) outliers, with the observed colors
    of the general population better tracked.  As can be seen, the model
    colors agree well with the raw data at $z\approx0.5$, $(r-i)\sim-0.1$.
    The model tracks begin to struggle to reproduce the observed colors at
    $z>2$, but remain consistent with the 1$\sigma$ dispersion until
    $z>2.5$.
    
    For $(i-z)$, just as for $(r-i)$, the model tracks replicate the
    mean colors very closely, all the way to $z\sim2.7$ and thus for the
    optical colors, the models perform best for these reddest bands.  At
    $z>2.7$ for $(i-z)$ and indeed all four optical colors, the models
    struggle to reproduce the observed quasar colors at these higher
    redshifts, consistently being too blue.

    \begin{figure*}
      \includegraphics[height=15.0cm,width=15.0cm] 
      {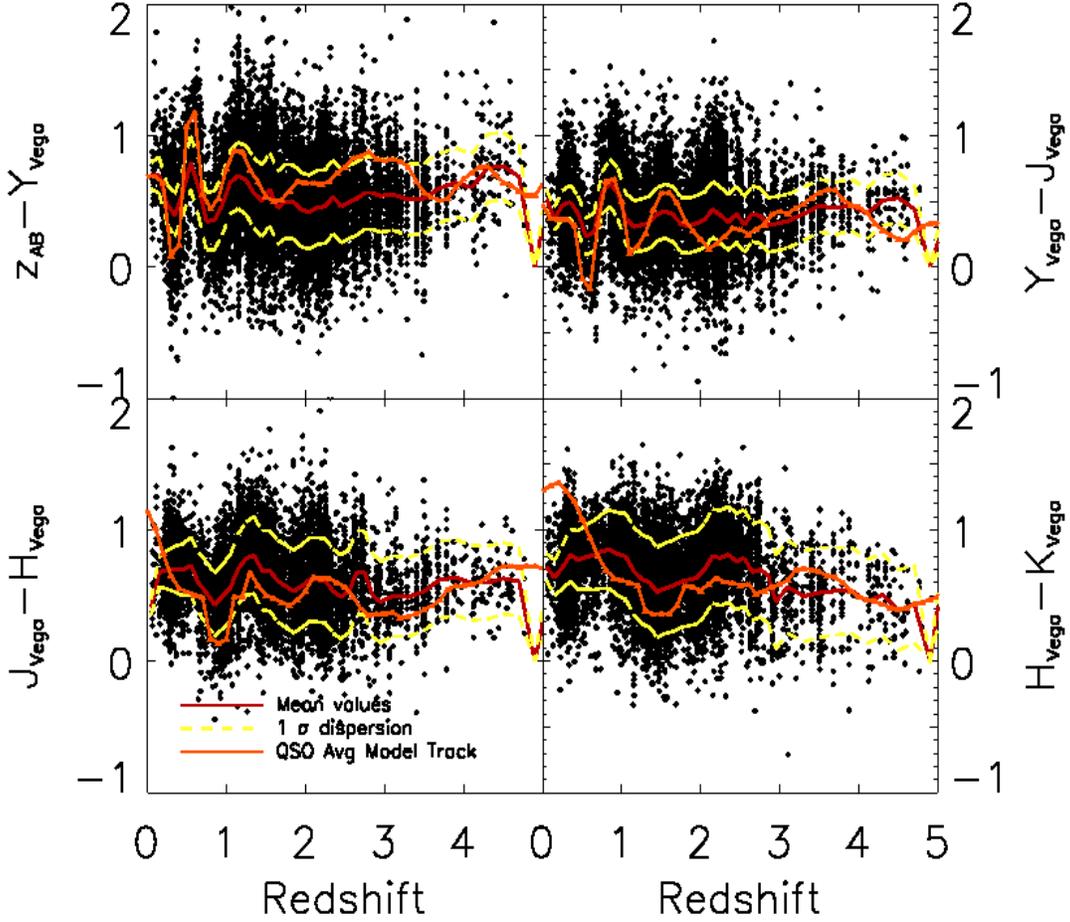} 
      \centering 
      \caption[Color versus redshift plots from Stripe 82 matches of $z-Y$,
        $Y-J$, $J-H$ and $H-K$.]
        {Color versus redshift plots from Stripe 82 matches. 
          $z-Y$ (top left); $Y-J$ (top right); $J-H$ (bottom left) and 
          $H-K$ (bottom right).  The given lines are the same as in
          Figure~\ref{fig:4col}. Note the dip and rise at $z\sim0.4, \sim0.7,
          \sim0.9$ and $\sim1.3$, for $z-Y$, $Y-J$, $J-H$ and $H-K$
          respectively, due to the location of the strong H$\alpha$ emission
          line in the NIR filters.} 
        \label{fig:4nircolors} 
    \end{figure*}    
    We tentatively suggest this is the same affect as has recently
    been reported in \citet{Prochaska09} and \citet{Worseck10}, namely,
    that the original SDSS color-selection that was used to select
    quasars, and from which the R09 and hence our own matched catalog is
    based, may be systematically biased as these redshifts. It has long
    been known that the redshifts around $z\sim2.7$ are very troublesome
    for selecting quasars, since the broadband colors are so similar to
    those of e.g. F5 V stars \citep[see e.g. Figure 1 of ][]{Fan99}, leading
    to very poor survey efficiency and heavy incompleteness
    \citep{Richards06, Croom09a}.  However, these recently analyses model
    in great detail the spectra of these high redshift objects, paying
    particular attention to the flux transmission (and decrement) blueward
    of Lyman-$\alpha$, in order to generate accurate mock UV and optical
    photometry.  Specifically, \citet{Worseck10} suggest that the SDSS
    color-selection selection systematically misses quasars with blue,
    $u-g\lesssim 2$ colors at $3\lesssim z \lesssim 3.5 $ due to the
    preference of selecting $z\gtrsim2.7$ quasars with intervening
    \ion{H}{1} Lyman limit systems (which generally turn quasars red).
    This would potentially begin to explain the offset between the model
    tracks and observed quasar colors in Figure~\ref{fig:4col}. We are
    keen to note however, that there are more recent versions of the
    quasar color models given in \citet{Hewett06} and that further
    analysis is required here.

    Noting the shape, response and wavelength coverage \citep[and
    indeed gaps between filters, see e.g. Figure 1][]{Hewett06} between the
    $Y$, $J$, $H$ and $K$ filters, we can now begin to explain the NIR
    color trends with redshift seen in Figure~\ref{fig:4nircolors}. The NIR
    color-redshift relations still exhibit structure, but arguably less
    than in the optical. This is mainly due to the fact that it is now the
    lines towards the redder end of the optical, and in particular
    H$\alpha$, that lead to the trends seen in Figure~\ref{fig:4nircolors}.
    We note that \citet{Assef10} and in particular \citet[][their table
    6]{Glikman06} are useful references for the following.
    
    The H$\alpha$ line is very much the dominant emission line over
    the rest-wavelength range, $\lambda=6500-21000$\AA. The blended
    \ion{He}{1} and Paschen-$\gamma$ lines at around 10,900\AA, the
    Paschen-$\beta$ line at 12,820\AA, and Paschen-$\alpha$ emission at
    18,756\AA, are also present but contribute little. For example, even
    if a Paschen line had an equivalent width of 100\AA, with the NIR
    bands being so broad (e.g. $\approx4000$\AA\, for the {\it K}-band; see
    Table~\ref{tab:filter_details}), the contribution here would be
    $\sim$a few \% at redshift $z=0$ - significantly smaller than the
    measurement errors in the colors. These emission lines are superimposed
    on the continuum power-law spectrum, $F({\nu})\propto \nu^{\alpha}$
    where $\alpha=-0.78$, measured over $\lambda=5700-10850$\AA\,
    \citep[][but see also \citet{Kishimoto08}]{Glikman06}.
    
    Thus, starting with the top left panel, $(z-Y)$, we note the dip
    at redshift $z\sim0.3$ is due to H$\alpha$ entering the $z$-band
    causing the general quasar color to be blue, while at redshift,
    $z\sim0.45$, H$\alpha$ is leaving $z$, and entering the $Y$-band
    causing the colors to become redder, reaching an observed mean (model)
    peak of $(z-Y)\approx 0.5$ (1.0) at redshift $z\approx0.6$.  At $z
    \sim 0.7$, H$\alpha$ leaves the $Y$-band.  The model tracks trace the
    observed colors well up to $z\sim 2$, but then grow progressively
    redder and redder at higher redshifts (though do remain within the
    measured 1$\sigma$ dispersion). This behavior, the model tracks being
    {\it redder} than the observed quasar colors at high, $z>2$, redshift is only
    see in $(z-Y)$.
    
    Moving to, $(Y-J)$, we again see a dip, at $z \sim 0.5$, towards
    bluer colors, as H$\alpha$ enters the $Y$-band, and then a rise
    towards redder colors, peaking at redshift $z \sim 0.9$ as H$\alpha$
    moves out of $Y$, and into the $J$-band (noting the gap between bands
    at 1.4$\mu$m). The models' continued rise and fall behavior at
    $z\sim1.1-2.1$ can be explained by H$\beta$, then \ion{O}{3} and then
    \ion{Mg}{2} marching through the WFCAM bands. The observed mean
    colors however, show a much smaller amplitude of color-change over
    this redshift range, though the individual objects seem to respond
    more to these emission lines.
    
    For $(J-H)$, the passing of H$\alpha$ first through the $J$-band
    and then through the $H$-band gives the trough at $z \sim 0.8-1.1$.
    Here the model tracks, even out to our maximum plotted redshift of
    $z=3$, replicate the observed quasar colors very well. The success of
    the model tracks in $(J-H)$ could potentially be used for quasar
    selection in future redshift surveys, including the implementation of
    ``KX'' completeness tests \citep[][]{Croom01KX, Smail08, Jurek08}.
    
    Finally, for $(H-K)$, the H$\alpha$ trough is across redshifts
    $z=1.2-1.8$. Note the broadening of the trough in the four NIR colors as
    H$\alpha$ passes to higher redshifts. Again the model tracks replicate
    the observed quasar colors well, though there is a large dispersion at
    all redshifts. The one place the models tracks are in disagreement is
    at low, $z<0.5$, redshifts. This is potentially due to flux from an
    underlying red host galaxy being over-represented, but more likely,
    the lack of data points to calculate the observed mean well.
    
    \begin{figure} [h]
      \includegraphics[height=9.0cm,width=9.0cm] 
      {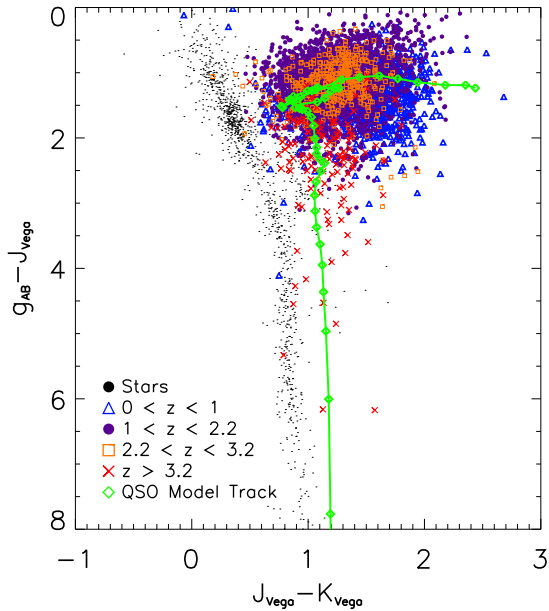}
      \centering 
      \caption[The location of matched quasars, and stellar sources in the
        $g-J$ vs. $J-K$ plane.] 
        {The location of matched quasars, and stellar sources in the
        $g-J$ vs. $J-K$ plane. Stars are solid (black) points; 
        $0<z<1$ quasars are open (blue) triangles; $1<z<2.2$ 
        quasars are solid (purple) circles; $2.2<z<3.2$ quasars
        are open (orange) squares and (red) crosses are $z>3.2$ 
        objects. The (``average'') model tracks of \citet{Hewett06}
        are given by the (green) line, with $z=0$ around $(J-K)=2.5$
        increasing in $\Delta z=0.1$ steps (green diamonds) to 
        $z=4.8$ at $(g-J)=8$. }
      \label{fig:gJK} 
    \end{figure}
    \subsection{Stellar Contamination}
   Following \citet{Maddox08} and \citet{Smail08}, we now investigate
    the location in $(g-J)$ vs. $(J-K)$ (``$gJK$'') color-space of our
    matched catalog.
    
    Here our motivation is not so much directed at creating complete
    ``KX'' selected samples, but is more towards determining if there is a
    certain part of color-space where quasars, and in particular mid,
    $z=2.2-3.5$, redshift quasars lie apart from stars. This is of key
    importance to two new spectroscopic surveys, the SDSS-III: Baryon
    Oscillation Spectroscopic Survey \citep[BOSS; ][]{Schlegel07} and the
    AAT-UKIDSS-SDSS (AUS) QSO Survey (P.I. S.M.~Croom).  Both the BOSS and
    AUS survey aim to gather spectroscopic information for $z>2.2$
    quasars, for investigations into cosmological parameters using the
    Lyman-$\alpha$ forest (BOSS) and global quasar population studies (AUS
    and BOSS). As such, any technique to reduce stellar contamination is
    of upmost interest to these survey teams. The KX technique is based
    upon the fact that at redshifts $z\approx=2-3$, NIR photometry samples
    the Rayleigh-Jeans tails of A and F stars (the major contaminants in
    an optical-based quasar selection), where flux is decreasing rapidly with 
    the increasing wavelength, whereas quasar SEDs are remaining relatively flat.
    
    In Figure~\ref{fig:gJK} we plot the $(g-J)$ vs. $(J-K)$ colors for
    the objects on Stripe 82 from our match catalog, with spectroscopic
    redshifts. We also select 5000 point sources lying on Stripe 82, from
    the SDSS CAS, with spectroscopic redshifts less than 0.02, and gather
    ULAS NIR photometry for these objects where available. These can be
    considered stars because of their low redshifts and morphology, and
    are solid (black) points in $gJK$ figure. $0<z<1$ quasars are open
    (blue) triangles; $1<z<2.2$ quasars are solid (purple) circles;
    $2.2<z<3.2$ quasars are open (orange) squares and (red) crosses are
    $z>3.2$ objects. Again we show the ``Average'' model tracks of
    \citet{Hewett06}, given by the (green) line, with $z=0$ around
    $(J-K)=2.5$ increasing in $\Delta z=0.1$ steps (green diamonds) to
    $z=4.8$ at $(g-J)=8$.

    \begin{figure} [h]
      \includegraphics[height=9.0cm,width=9.2cm] 
      {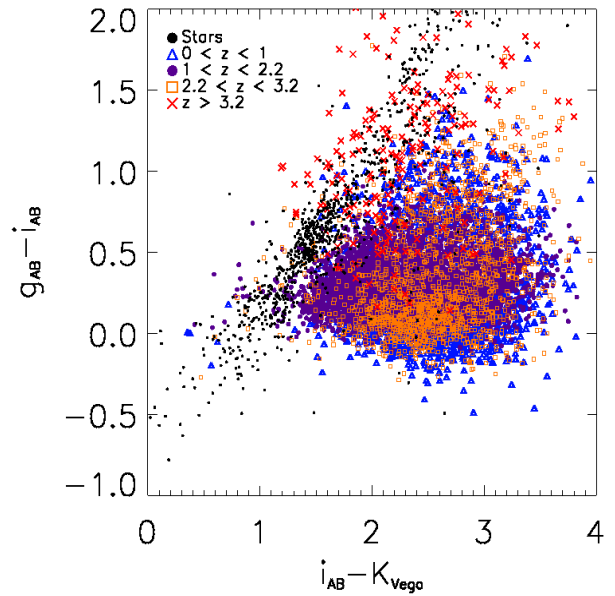}
      \centering 
      \caption[] 
      {$g-i$ vs. $i-K$ diagram to show the colors of Stripe 82 objects 
        from the matched catalog.  Stars are solid (orange) points.
        Quasars are displayed in bins of varying redshift ranges: $z<1.0$, black stars;
        $1.0<z<2.2$ blue open triangles; $2.2<z<3.2$ orange open squares, and
        greater than 3.2, red crosses. } 
      \label{fig:gik3d} 
    \end{figure}
    
    Our results are qualitatively very similar to that of
    \citet{Maddox08}, that is, ({\it a}) the stellar locus traces out a
    clear band in $gJK$ color-space, related to the different stellar
    spectral type and ({\it b}) that in general, stars appear to have
    $gJK$ colors that are in a distinct region of color space than
    quasars, especially those with $2.2<z<3.2$ (orange) open squares in
    Figure~\ref{fig:gJK}.  We find this very encouraging and suggest that
    selection in the $gJK$ color-plane could be of good utility to the
    BOSS and AUS surveys. We also note the almost complete lack of quasars
    in our $gJK$ plot that have colors similar to those of (model)
    galaxies around $(J-K)\approx1.7$ and $(g-J)\approx4.5$.  However, a
    strong caveat that has to be imposed here is to recall that the
    original R09 quasar selection was predominantly geared towards $z<2$
    ``UVX'' selected, non-extended objects.  This is to be compared with
    the study by \citet{Smail08} who find that lower, $z\sim0.45$ compact
    galaxies are the main and major contaminent of their ``KX''
    selection. Interestingly, \citet{Smail08} report that the spectral mix
    of the contaminating population includes sources whose spectra are
    best fit by both broadline and narrow-line AGN templates, as well as
    narrow emission-line, absorption-line and spiral galaxy templates.
    Indeed, early analysis from the BOSS (Ross et al., 2010, in prep.) suggests that
    these morphologically compact, narrow-emission line galaxies at lower,
    $z\approx0.2-0.5$ redshift, continue to be selected as $z>2.2$
    quasars.
    
    Inspired by \citet{Mehta10}, we also plot the location of Stripe
    82 objects from the matched catalog, this time in $g-i$ vs. $i-K$
    ($giK$) color-space, see Figure~\ref{fig:gik3d}. Again, quasars are
    displayed in bins of varying redshift ranges: $z<1.0$, black stars;
    $1.0<z<2.2$ blue open triangles; $2.2<z<3.2$ orange open squares, and
    greater than 3.2, red crosses.  On initial inspection, we again see
    the stellar locus inhabiting a distinct sequence in color-space, which
    is somewhat separate to that of the quasars. However, we leave further
    investigations into the use of the $(g-i)$ vs. $(i-K)$ color-space for
    star-quasar-galaxy identification, and survey completeness to future
    study.

\section{High Redshift Quasars}\label{sec:highz}
    
    \subsection{High Redshift objects in Matched Catalog}
\begin{figure}
        \includegraphics[height=8.0cm,width=10.0cm] {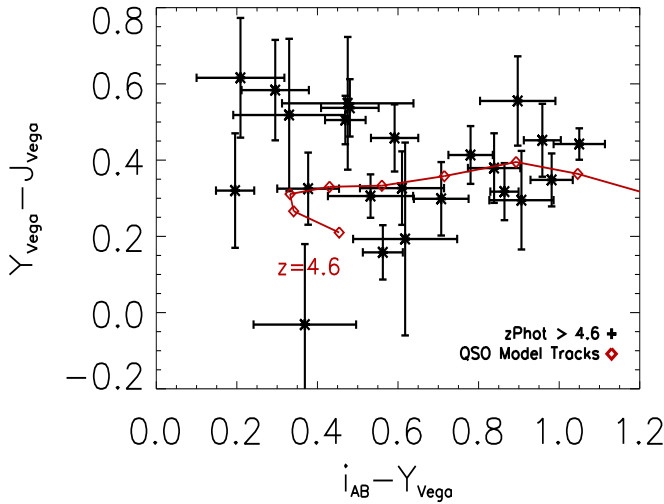}
        \centering
        \caption[] {High, $4.6<z<5.35$ redshift quasars in $i-Y$ vs
          $Y-J$ color space.  The ``average'' quasar model tracks of
          \citet{Hewett06} are again given, starting at $z=4.6$ (labelled) and
          then increasing in $\Delta z=0.1$ steps, shown by the open (red)
          diamonds.}
        \label{fig:iyj}
      \end{figure}

    Figure~\ref{fig:iyj} plots high-redshift quasars, those with $z
    \gtrsim 4.6$ (essentially $r$-band drop-outs), in $iYJ$ color-color
    space. Only objects with both a ``good'' photometric and
    spectroscopic redshifts are plotted. For a photometric redshift to be
    considered ``good'' it must be within the error limits of the
    spectroscopic redshift.  Out of 49 high-redshift ($z \gtrsim 4.6$)
    quasars possessing both a photometric and spectroscopic redshift, 40
    have ``good'' photometric redshifts.  Out of this group of 40 quasars,
    23 objects possess both $Y$ and $J$ magnitudes, and we plot these in
    Figure \ref{fig:iyj}. These objects span the redshift range of
    $4.606 \leq z_{\rm spec} \leq 5.289$. 
    No obvious trends are visible, arguably due to
    our small sample size. However, we can say that these high-redshift
    quasars have roughly the same color and follow the given high-$z$
    model tracks of \citet{Hewett06} well.  

    \subsection{Very High Redshift objects}\label{sec:vhighz}
    \begin{table*}
      \begin{center}
          \caption{Photometric properties of 6 very high $z>5.7$, redshift
            quasars as detected in the ULAS DR3, along with their spectroscopic
            redshifts.  Positions are in degrees and J2000 coordinates.
            References: $^{a}$Venemans et al. (2007). $^{b}$Jiang et
            al. (2008). $^{c}$Fan et al. (2006). $^{d}$Mortlock et al. (2009). }
            \setlength{\tabcolsep}{2pt}
          \begin{tabular}{cc c ccccc cccc}
          \hline 
          \hline 
          R.A. & Decl. & redshift              & $u$                          & $g$                          & $r$                          & $i$                           & $z$                          & $Y$      & $J$ & $H$ & $K$ \\ 
          \hline
          $^{a,b}$30.8849   &   0.2081 & 5.850$\pm$0.003 & --                        & --                       & --                         & 23.72$\pm$0.22 & 20.87$\pm$0.10 & 19.85$\pm$0.12  & 19.06$\pm$0.10  & 17.76$\pm$0.07 & 17.32$\pm$0.09 \\
          $^{b}$58.45719 &   1.0680 & 6.049$\pm$0.004 & --                        & --                       & --                         & 24.03$\pm$0.30 & 20.54$\pm$0.08 & 20.12$\pm$0.16  & 19.46$\pm$0.16  & 18.55$\pm$0.17 & 18.16$\pm$0.22 \\
          $^{c}$129.1827 &   0.9148 & 5.82$\pm$0.02     & 26.21$\pm$0.57 & 25.80$\pm$0.55 & 22.75$\pm$0.23  & 21.23$\pm$0.09 & 18.86$\pm$0.05 & 18.26$\pm$0.03 & 17.70$\pm$0.029 & 17.03$\pm$0.03 & 16.19$\pm$0.03 \\
          $^{d}$199.7970 &   9.8476 & 6.127$\pm$0.004 & --                       & --                        & --                         & 22.55$\pm$0.09 & 19.99$\pm$0.03 & 19.22$\pm$0.06 & 18.69$\pm$0.05   & --       & 17.55$\pm$0.08 \\
          $^{c}$212.7970 & 12.2937 & 5.93$\pm$0.02     & 24.77$\pm$0.99 & 24.79$\pm$0.71 & 23.22$\pm$0.28  & 22.86$\pm$0.30 & 19.65$\pm$0.08 & 19.47$\pm$0.07 & 19.22$\pm$0.093 & 18.29$\pm$0.09 & 17.46$\pm$0.08 \\
          $^{c}$245.8826 & 31.2001 & 6.22$\pm$0.02     & 23.96$\pm$0.91 & 25.44$\pm$0.62 & 24.88$\pm$ 0.58 & 24.50$\pm$0.62 & 20.07$\pm$0.10 & --                        & 19.17$\pm$0.111 & --  & -- \\
            \hline
            \hline
            \label{tab:fan}
          \end{tabular}
       \end{center}
      \end{table*}

      For our final investigations, we actually leave the matched
      quasar catalog, and concentrate solely on the ULAS DR3 to investigate
      the NIR photometric properties of the very high, $z>5.7$, redshift
      objects that are given in the series of Fan et al., and Jiang et
      al. papers papers, \citep[][respectively and references
      therein]{Fan06, Jiang08, Jiang09}.  We also note the 19 very high
      redshift quasars discovered by the Canada-France High-$z$ Quasar
      Survey \citep[CFHQS; ][]{Willott07, Willott09, Willott10}. However,
      the CFHQS is not covered by the ULAS DR3, and only has the one ($J$)
      out of the 4 ULAS bands, so the NIR properties of these objects are
      not reported here.
      
      Out the 30 objects - 19 from the Fan et al. studies, 11 from
      Jiang et al.  with one object in common and the additional object from
      \citet{Mortlock09}, 13 (12) are within the ULAS DR3 ($K$-band)
      footprint, and 6 were detected in one or more of the ULAS bands.  The
      photometric properties of these 6 detections, along with their
      spectroscopic redshifts, are given in Table~\ref{tab:fan}.  5 of the 6
      objects have $K$-band detections, with magnitudes ranging from $K=
      16.19 - 18.16$. The reason for only $5/12$ having $K$-band detections
      we suggest is primarily due to the limiting magnitude for the ULAS. As
      a quick check, we note that the Jiang et al. objects were discovered
      using the deeper SDSS optical imaging, this means, potentially 6 out
      of the 7 $K$-band non-detections are fainter than the ULAS $K$-band
      limit. The mean $J$-band magnitude from the 6 objects presented in
      \citet{Jiang09} is $J\approx21.0\pm0.2$, whereas for the objects in
      Table~\ref{tab:fan} it is $J\approx18.9\pm0.3$. Thus if the colors of
      these $z\sim6$ objects are $(J-K)\sim1.1-1.8$, even the brightest and
      reddest object in \citet{Jiang09} would be right on the ULAS $K$-band
      limit, with the average object being closer to $K=19.3$.
      
      Our results are generally consistent with \citet{Venemans07} who
      report the discovery of the quasar at $z = 5.85$ (designated ULAS
      J020332.38+001229.2) after analysis of 106 deg$^{2}$ of sky from
      UKIDSS DR1, and \citet{Mortlock09} who report the discovery of ULAS
      J1319+0950, a quasar at $z = 6.13$. \citet{Glikman08} surveyed 27.3
      deg$^{2}$ of the ULAS EDR, but found no $z>6$ objects from
      spectroscopy of a 34 candidate list.
      
      Deeper and more complete surveys will be needed to get a better
      estimate on the spatial density of high-redshift quasars. To this
      affect, the Visible and Infrared Survey Telescope for Astronomy
      \citep[VISTA; ][]{Emerson04, Emerson10} has recently been commissioned,
      and is currently ramping up its suite of six imaging surveys
      \citep{Arnaboldi10}. Of direct interest to the high redshift Universe
      will be the ``VISTA Hemisphere Survey'', (VHS; PI: R.G. McMahon),
      which aims to image the entire \hbox{$\sim$20 000deg$^{2}$} Southern
      Sky, down to $J_{\rm AB} =21.2$ and $Ks_{\rm AB}=20.0$ ($J\approx20.3$
      and $K\approx18.1$).  On a longer timescale, the Synoptic All-Sky
      Infrared \citep[SASIR; ][]{Bloom09} is planned. The hope and goals of
      these surveys will be to find 100's (VISTA) or potentially 1000's
      (SASIR) of $z>6$ quasars. Whether these numbers are borne out in the
      new observations, will, of course, be of great future interest.
      
\section{Conclusions}
We have matched the optical, photometrically selected quasar catalog
of R09 to detections from the ULAS DR3.  The positions of the
1,015,082 objects from the DR6pQ catalog were uploaded to the WFCAM
Science Archive. To count as a match, the returned NIR object must
have been the closest object in the ULAS to the SDSS coordinates,
selected from within a pairing radius of 1$''$.  As long as at least 1
NIR band contained a non-default, i.e. not -9.9999$\times 10^{8}$
value, the object was considered a good match.

Our final catalog has \hbox{130,827} objects with detections in the
optical and one or more NIR bands; of which \hbox{74,351} objects have
{\it K}-band detections and \hbox{42,133} objects have the full 9-band
photometry.  Using this matched catalog, we present the following main
conclusions:

\begin{itemize}
\item{The positional standard deviation of the SDSS Quasar to ULAS 
    matches is $\delta_{\rm R.A.} = 0.1370''$ and $\delta_{\rm Decl.}=
    0.1314''$.  We find an absolute systematic astrometric offset between
    the SDSS Quasar catalog and the ULAS, of $|{\rm R.A._{offset}}| =
    0.025''$, and $|{\rm Decl._{offset}}| = 0.040''$; we suggest the nature
    of this offset to be due to the matching of catalogs, rather than
    image level data.}
\item{Our matched catalog has a surface density of $\sim$108
    deg$^{-2}$, for objects detected in any of the four NIR bands, and
    $\approx53$ deg$^{-2}$ for objects $K\leq18.27$. This compares to the
    $\approx122$ deg$^{-2}$ found from the R09 DR6pQ catalog and $85-150$
    deg$^{-2}$ down to $K\leq 20$ from \citet{Smail08}.}
\item{Tests using our matched catalog, along with data from the UKIDSS
    DXS, implies that our limiting magnitude is $i\approx20.6$ and that
    the reddest, $(r-K) \geq 5.0$ objects turn out to be either associated
    with, or contaminated by, a foreground extended source.}
\item{We plot redshift histograms for the Stripe 82 subsample,
    $K$-matches sample and total matched sample. The photometric
    $N(z\sim2.3)$ spike seen for the ``total sky'' sample appears to be
    dramatically reduced (although does not disappear completely) for the
    {\it K}-band matches. The photometric and spectroscopic histograms now
    seem to be in reasonable agreement, though there are possible
    deficiencies of photometric objects at $z\sim0.7$ and in particular,
    $z\sim1.9$.}
\item{Color-redshift diagrams, for the optical and NIR, show the close
    agreement between our matched catalog and the ``average'' models of
    \citet{Hewett06}, at redshift $z\lesssim2.0$. At higher redshifts, the
    models generally appear to be generally bluer than the mean observed
    quasar colors.  We {\it tentatively} suggest this is the same affect
    as has recently been reported in \citet[][]{Worseck10}, namely, that
    the original SDSS color-selection that was used to select quasars (and
    ultimately our own matched catalog) may be systematically biased
    towards missing blue $(u-g)\leq2.0$ quasars at $z\sim2.2-3.5$, due to
    the selection preferentially selecting intervening \ion{H}{1} Lyman
    limit systems.}
\item{Since stellar contamination is of great interest for ongoing
    quasar surveys such as the SDSS-III:BOSS and AUS, we plot a test 
    set of stellar data with our matched catalog data in $gJK$ and $giK$ 
    color space. We confirm findings from previous studies, in particular 
    \citet{Maddox08}, that $(a)$ the stellar locus traces
    out a clear band in $gJK$ color-space, related to the different
    stellar spectral type and $(b$ that in general, quasars, 
    especially those with $2.2 < z < 3.2$ mostly lie
    in a distinct region of $gJK$ color space than stars.}
\item{We plot the $iYJ$ colors of our matched catalog for
    high, $z>4.6$, redshift spectroscopically confirmed quasars, again 
    comparing to model tracks, though no obvious trends 
    are seen. Finally, just using the ULAS DR3 and the very high, $z>5.7$,
    redshift objects reported in \citet{Fan06}, \citet{Jiang08}, \citet{Jiang09} 
    and \citet{Mortlock09}, we find 6 (5) out of 13 (12) quasars have 
    NIR ($K$) band detections.}
\end{itemize}

It is worthwhile to mention that many of our tests have also been
performed very recently in \citet{Wu10}, and although we have not
performed any direct comparisons between the results found herein and
their study, we see very good general agreement, e.g. for color-$z$
relations, between the two investigations.

The major motivation for the construction of the matched catalog
presented here, was its utilization in a future study where we measure
the $K$-band quasar luminosity function.  As quasars are measured
further in the NIR, the flux due to host galaxies is no longer
negligible but will rather constitute a sizeable percentage of the
total bolometric flux from a quasar.  Our future work (Peth, Ross et
al. in prep.), will address this, and other issues in order to
construct and measure the observed $K$-band quasar luminosity
function.

Advancement can come from technological breakthroughs as well as new
theoretical insights. Future telescopes and surveys, e.g. VISTA-VHS
and the SASIR, will not only cover more of the sky but should also be
able to observe to greater depths.  Future observations will warrant
an updated analysis of quasar properties in multiple bands, and will
be both necessary and essential to further understand the formation
and evolution of quasars.

\section*{Acknowledgments}
This work was supported by National Science Foundation grants
AST-0607634 (M.A.P., N.P.R. and D.P.S.).  We warmly thank M.A. Read
for providing the matched catalogs. R.G. McMahon provided very kind
input and information regarding the ULAS, especially for the
discussions regarding the behaviour of the magnitude errors. P.Hewett,
G.T. Richards and J.P. Stott provided useful discussion and comments.
We thank the referee for a timely report that has improved our
manuscript, and we thank The Astronomical Journal for an extension to
the deadline for the submission of our revisions.

The JavaScript Cosmology Calculator was used whilst preparing this
paper \citep{Wright06}. This research made extensive use of the NASA
Astrophysics Data System. The data will become publicly available upon
publication of this paper.

Funding for the SDSS and SDSS-II has been provided by the Alfred
P. Sloan Foundation, the Participating Institutions, the National
Science Foundation, the U.S. Department of Energy, the National
Aeronautics and Space Administration, the Japanese Monbukagakusho, the
Max Planck Society, and the Higher Education Funding Council for
England. The SDSS Web Site is http://www.sdss.org/.

The SDSS is managed by the Astrophysical Research Consortium for the
Participating Institutions. The Participating Institutions are the
American Museum of Natural History, Astrophysical Institute Potsdam,
University of Basel, University of Cambridge, Case Western Reserve
University, University of Chicago, Drexel University, Fermilab, the
Institute for Advanced Study, the Japan Participation Group, Johns
Hopkins University, the Joint Institute for Nuclear Astrophysics, the
Kavli Institute for Particle Astrophysics and Cosmology, the Korean
Scientist Group, the Chinese Academy of Sciences (LAMOST), Los Alamos
National Laboratory, the Max-Planck-Institute for Astronomy (MPIA),
the Max-Planck-Institute for Astrophysics (MPA), New Mexico State
University, Ohio State University, University of Pittsburgh,
University of Portsmouth, Princeton University, the United States
Naval Observatory, and the University of Washington.

\appendix
\section{A. Photometric Bands and Conversions}
    Due to the differing normalizations between the
    SDSS and  UKIDSS photometric systems, certain corrections are required.  To present
    our data in the  purest sense, all the NIR magnitudes from UKIDSS
    (originally AB magnitudes)  were corrected to Vega magnitudes as
    suggested in \citet{Hewett06}.
    
    Although ULAS magnitudes are reported in terms of Vega and SDSS
    magnitudes are reported in AB terms for the most part whenever an
    optical-NIR color was calculated both magnitudes were left in their
    default term.
    
\begin{table}[h*]
  \begin{center}
   \caption{Adapted from Table 19 of \citet{Stoughton02}, 
      Table 1 of \citet{Lawrence07} and Table 7 of
      \citet{Hewett06}.}
    \setlength{\tabcolsep}{4pt}
    \begin{tabular}{lcllcl}
      \hline
      \hline
      Band & $\lambda_{\rm eff}  {\buildrel _{\circ} \over {\mathrm{A}}}$ 
              &  $\lambda_{\rm min} {\buildrel _{\circ} \over {\mathrm{A}}}$ 
              & $\lambda_{\rm max} {\buildrel _{\circ} \over {\mathrm{A}}}$ 
              & FWHM ${\buildrel _{\circ} \over {\mathrm{A}}}$ 
              & AB - Vega Transformations \\
      \hline
      {\it u} & 3551  &  3005 &  4000 &  581 & $u$ = $u_{AB}$ - 0.927 \\
      {\it g} & 4686  &  3720 &  5680 & 1262 & $g$ = $g_{AB}$ + 0.103       \\
      {\it r} & 6166  &  5370 &  7120 & 1149 & $r$ = $r_{AB}$ - 0.146  \\
      {\it i} & 7480  &  6770 &  8380 & 1237 & $i$ = $i_{AB}$ - 0.366 \\
      {\it z} & 8932  &  8000 & 10620 &  994 & $z$ = $z_{AB}$ - 0.533  \\
      {\it Y} & 10305 &  9790 & 10810 & 1020 & $Y$ = $Y_{AB}$  - 0.634           \\
      {\it J} & 12483 & 11690 & 13280 & 1590 & $J$  = $J_{AB}$- 0.938           \\
      {\it H} & 16313 & 14920 & 17840 & 2920 & $H$ = $H_{AB}$ - 1.379          \\
      {\it K} & 22010 & 20290 & 23800 & 3510 & $K$ = $K_{AB}$ - 1.9            \\ 
      \hline
      \hline
      \label{tab:filter_details}
    \end{tabular}
     \end{center}
\end{table}


\bibliographystyle{mn2e}
\bibliography{tester_mnras}

\end{document}